\documentclass[prx,letterpaper,aps,superscriptaddress,floatfix,twocolumn,nofootinbib]{revtex4-1}
\pdfoutput=1
\usepackage{graphicx}
\usepackage{amsmath}
\usepackage{amsfonts}
\usepackage[normalem]{ulem}
\usepackage[small,bf]{subfigure}
\usepackage[utf8]{inputenc}
\usepackage{dsfont}
\usepackage{xcolor}
\usepackage{slashed}
\newcommand{\beq}{\begin{equation}}
\newcommand{\eeq}{\end{equation}}
\newcommand{\bk}{{{\bf{k}}}}
\newcommand{\bs}{{{\bf{s}}}}

\newcommand{\br}{{{\bf{r}}}}

\newcommand{\beqa}{\begin{eqnarray}}
\newcommand{\eeqa}{\end{eqnarray}}
\newcommand{\pdg}{{\vphantom \dag}}
\newcommand{\dg}{{\dag}}
 
\newcommand{\bsigma}{{\boldsymbol \sigma}}

\newcommand{\upa}{\uparrow}
\newcommand{\da}{\downarrow} 
\newcommand{\ra}{\rightarrow} 

\newcommand{\cH}{{\cal H}}

\newcommand{\Z}{\mathbb{Z}}
\begin{document}
\title{Unquantized anomalies in topological semimetals}
\author{L. Gioia}
\affiliation{Department of Physics and Astronomy, University of Waterloo, Waterloo, Ontario 
N2L 3G1, Canada} 
\affiliation{Perimeter Institute for Theoretical Physics, Waterloo, Ontario N2L 2Y5, Canada}
\author{Chong Wang}
\affiliation{Perimeter Institute for Theoretical Physics, Waterloo, Ontario N2L 2Y5, Canada}
\author{A.A. Burkov}
\affiliation{Department of Physics and Astronomy, University of Waterloo, Waterloo, Ontario 
N2L 3G1, Canada} 
\affiliation{Perimeter Institute for Theoretical Physics, Waterloo, Ontario N2L 2Y5, Canada}
\date{April 2, 2021}
\begin{abstract}
Topological semimetals are a new class of metallic materials, which exist at band fillings that ordinarily 
correspond to insulators or compensated accidental semimetals with zero Luttinger volume. 
Their metallicity is a result of nontrivial topology in momentum space and crystal symmetry, wherein 
topological charges may be assigned to point band-touching nodes, preventing gap opening, unless protecting 
crystal symmetries are violated. 
These topological charges, however, are defined from noninteracting band eigenstates, which raises the possibility that 
the physics of topological semimetals may be modified qualitatively by electron-electron interactions. 
Here we ask the following question: what makes the topological semimetals nontrivial beyond band theory? Alternatively, can strong electron-electron interactions open a gap in topological semimetals 
without breaking the protecting symmetries or introducing topological order? We demonstrate that the answer is generally no, and 
what prevents it is their topological response, or quantum anomalies. While this is familiar in the case of magnetic Weyl semimetals, where the topological response 
takes the form of an anomalous Hall effect, analogous responses in other types of topological semimetals are more subtle and involve crystal symmetry as well as electromagnetic gauge fields. Physically these responses are detectable as fractional symmetry charges induced on certain gauge defects. We discuss the cases of type-I Dirac semimetals and time-reversal invariant Weyl semimetals in detail. For type-I Dirac semimetals, we also show that the anomaly vanishes, in a nontrivial manner, if the momenta of the Dirac nodes satisfy certain exceptional conditions.
\end{abstract}
\maketitle

\section{Introduction}
\label{sec:1}
While the concepts of nontrivial electronic structure topology have traditionally been associated with insulators~\cite{Hasan10,Qi11}, recent 
work has lead to the realization that gapless metallic states may also be topological~\cite{Volovik03,Volovik07,Weyl_RMP}. 
According to the standard band theory of crystalline solids, whether a given material is a metal or an insulator is determined by the electron filling per unit cell. 
When the filling is an odd integer, we necessarily get a metal with a Fermi surface, whose volume is directly determined by the filling and is not renormalized by the 
electron-electron interactions~\cite{Luttinger60,Oshikawa00,Hastings04}. 
When the filling is an even integer, on the other hand, the net Fermi surface volume must be zero, which corresponds to either an insulator or a compensated semimetal, 
with electron and hole Fermi surfaces enclosing equal volume. The compensated semimetal arises due to fact that the bands may overlap in energy and is accidental, 
in the sense that the overlap may be removed without altering the crystal symmetry, whether it exists or not is a matter of microscopic detail.

Recently discovered topological semimetals are different from compensated semimetals in that their existence is not accidental, they arise inevitably under certain 
conditions, either as intermediate phases between topologically distinct insulators or in crystals with certain symmetry groups. 
In particular, Weyl semimetal (WSM) arises as either an intermediate phase between a quantum anomalous Hall insulator and an ordinary three-dimensional (3D) insulator~\cite{Burkov11-1};
or as an intermediate phase between a time reversal (TR) invariant 3D topological insulator and an ordinary 3D insulator, when inversion symmetry is violated~\cite{Murakami07}. 

When TR and inversion symmetry are present and all bands are thus doubly degenerate, pairs of opposite-chirality Weyl nodes must occur at the same momenta in the
Brillouin zone (BZ), which generally means that a gap is opened. 
However, certain point group crystal symmetries may protect four-fold degenerate band-touching points. Such materials are called Dirac semimetals (DSM). 
These come in two classes, type-I and type-II~\cite{Nagaosa14}. 
In type-I Dirac semimetals, such as Na$_3$Bi and Cd$_3$As$_2$~\cite{Fang12,Fang13}, Dirac points occur in pairs at generic BZ momenta on an axis of rotation, and are protected by 
a symmetry of rotations about this axis. This type of Dirac semimetal arises as an intermediate 
phase between an ordinary insulator and a weak topological insulator, in which the direction of the weak index (a reciprocal lattice vector) coincides with the rotation axis. 
In type-II Dirac semimetals, in contrast, there is a single Dirac node at a time reversal invariant momentum (TRIM) at the edge of the BZ, terminating an axis of nonsymmorphic 
rotation~\cite{Kane12,Steinberg14} (the minimum total number of such Dirac points in the BZ is still two, unless TR is explicitly broken). 
Such a Dirac semimetal is not an intermediate phase between two insulators, but exists in crystals with certain nonsymmorphic symmetry groups, which inevitably have 
four-fold band degeneracies at TRIM at the edge of the BZ~\cite{Nagaosa14,Parameswaran13,Parameswaran19}. 

The band-touching points in both Weyl and Dirac semimetals are stable as long as the protecting symmetries are present or as long as the points are not 
pairwise annihilated by bringing them to the same position in the BZ (this applies to Weyl and type-I Dirac semimetals). 
This stability may be connected with the existence of a momentum-space topological invariant, associated with the band-touching point. 
In the case of Weyl semimetals, this topological invariant is a nonzero Chern number ($\pm 1$) of any closed surface in momentum space, enclosing 
the node. In the Dirac semimetal case, the invariant is more subtle and involves counting rotation eigenvalues of occupied and empty states on the rotation 
axis on the opposite sides of the Dirac point~\cite{Furusaki15}. 

An important limitation of this picture is that it is based on noninteracting band eigenstates. A question then arises to what extent topological semimetals are stable 
with respect to the electron-electron interactions. 
By stability here we do not mean perturbative stability with respect to gap opening: all 3D point-node semimetals are stable in this sense thanks to the vanishing density 
of states at the Fermi energy. Rather, we are interested in the question to what extent their topologically nontrivial nature is still manifest when the interactions are not weak. 
We have recently addressed this issue in the simplest case of a magnetic Weyl semimetal~\cite{Wang20,Thakurathi20,Sehayek20} (see Refs.~\cite{Meng16,Morimoto16,Sagi18,Meng19,Teo19} for related work). 
In this paper we generalize this earlier work to TR-invariant Weyl and type-I Dirac semimetals. As explained above, what unifies these three classes 
of topological semimetals is that they arise as intermediate phases between topologically-distinct insulators. 
Type-II Dirac semimetals are not of this type, leading to a significantly different physics, which we will address in a separate publication. 

A way to formulate this question precisely in the case of a magnetic Weyl semimetal is as follows. 
In addition to topological invariants, formulated in terms of band eigenstates, magnetic Weyl semimetals are also characterized 
by topological response, which, in particular, takes the form of an anomalous Hall effect~\cite{Burkov11-1}. 
Specializing to the simplest case of a Weyl semimetal with a single pair of opposite-chirality nodes, the anomalous Hall conductivity is proportional to the distance between 
the nodes in momentum space
\beq
\label{eq:AHC}
\sigma_{xy} = \frac{1}{2\pi} \frac{2Q}{2 \pi}\quad,
\eeq
where we are using $\hbar = e = c = 1$ units here and throughout this paper and the nodes are taken to be located at $k_z = \pm Q$. 
This Hall conductivity, which is a fraction of a conductivity quantum $1/2 \pi$ per atomic layer, is a characteristic property of magnetic Weyl semimetals,
which is well-defined even when the band eigenstates are not. 
We may then ask whether a trivial gapped insulator at the same electron filling per unit cell as the Weyl semimetal may have the Hall conductivity given 
by Eq.~\eqref{eq:AHC}. 
The answer to this is no since Eq.~\eqref{eq:AHC} implies a $\frac{\sigma_{xy}}{2} A dA$ term in the Lagrangian for the electromagnetic field, which is not invariant with respect to large gauge transformations. 
Gapless modes are needed to restore gauge invariance, which, in the absence of a Fermi surface, makes the Weyl nodes necessary. 
However, a fractionalized insulator with a particular type of topological order is consistent with Eq.~\eqref{eq:AHC} when $2Q = \pi$, taking the lattice constant in the $z$-direction to be unity~\cite{Wang20}. 

Here we ask whether this line of reasoning may be generalised to other point-node topological semimetals, in particular TR-invariant Weyl and type-I Dirac semimetals. 
It is not at all obvious that this is possible since, unlike the magnetic Weyl semimetal, these do not possess any topological electromagnetic responses.
This makes one wonder if the nontrivial topology of TR-invariant Weyl and type-I Dirac semimetals only exists in the weakly interacting limit. 
In this paper we demonstrate that this is not the case. 
We show that both TR-invariant Weyl and type-I Dirac semimetals possess ``unquantized" topological responses, similar to the magnetic Weyl semimetal,
except involving crystalline symmetry, rather than purely electromagnetic, gauge fields. 
These manifest as fractional electric charge density induced on crystalline symmetry defect (i.e. dislocations and disclinations) configurations.

Alternatively, the Hall conductivity of a magnetic Weyl semimetal Eq.~\eqref{eq:AHC} may be viewed as being a consequence of a nonzero charge density, induced in the ground state 
of the Weyl semimetal by an applied magnetic field, given by the Streda formula
\beq
\label{eq:Streda}
\sigma_{xy} = \left(\frac{\partial n}{\partial B} \right)_{\mu}\quad. 
\eeq
Similarly, we demonstrate that topological responses of TR-invariant Weyl and type-I Dirac semimetals may be expressed in terms of nontrivial ground state symmetry charges,
induced by the applied magnetic field. These symmetry charges are the crystal momentum (translational symmetry charge) in the case of the TR-invariant Weyl semimetal and 
the angular momentum (rotational symmetry charge) in the case of the type-I Dirac semimetal.

The rest of the paper is organized as follows. 
In Section~\ref{sec:prelim} we introduce and review the concepts of unquantized anomalies and of symmetry gauge fields, which may be unfamiliar to some readers. 
In Section~\ref{sec:1+1d} we discuss a series of one-dimensional lattice models, which introduce the mixed crystalline symmetry-electromagnetic anomalies in the simplest 
possible setting. We show that these anomalies may be viewed as a generalization of the familiar notion of fractional $U(1)$ charge density, which is formally related to the $(1+1)d$ chiral anomaly, to discrete symmetry gauge fields. 
In Section~\ref{sec:2} we generalize the results of Section~\ref{sec:1+1d} to 3D topological semimetals. As familiar from the standard discussions of the chiral anomaly, 
when the semimetals are placed in an external magnetic field, the resulting lowest Landau levels encode the anomaly physics and connect the anomalies in $3+1$ dimensions to their 
$1+1$-dimensional counterparts. 
Finally, we generalize the vortex condensation analysis of Ref.~\cite{Wang20} to the cases of TR-invariant Weyl and type-I Dirac semimetal, 
which provides yet another viewpoint on their topological nontriviality in the presence of strong interactions. 
We conclude in Section~\ref{sec:conclusion} with a brief summary and discussion of the main results of the paper. 

\section{Preliminaries}
\label{sec:prelim}
Since we will be using a number of concepts, such as anomalies and symmetry gauge fields, that may be unfamiliar to some readers, in this section we will briefly review these concepts.

\subsection{Topological response and ``unquantized quantum anomalies''}
We will start by reviewing how certain types of unquantized topological response can constrain the low energy phases in a way that is similar to the usual quantum anomalies. 

We illustrate the idea using a familiar example. Consider a $(2+1)d$ fermion system with charge $U(1)$ symmetry. If the Hall conductance $\sigma_{xy}=-\sigma_{yx}$ is not an integer (in units of $e^2/h=1/2\pi$), then it is well known that the ground state cannot be short-range entangled: if the ground state is gapped, it should realize a non-trivial topological order, as in fractional quantum Hall effects; the ground state can also be gapless, for example by having a Fermi surface that encloses a Berry phase $\Phi=4\pi^2\sigma_{xy}$~\cite{Haldane04}. One way to see this is to notice that if the ground state is short-range entangled, the theory of response to a probe $U(1)$ gauge field should be expressed as the integral of a local term. The Hall conductance corresponds to the familiar Chern-Simons (CS) term:
\beq
\label{eq:CS}
\int d^3x \frac{k}{4\pi}AdA\quad,
\eeq
where $\sigma_{xy}=k/2\pi$. It is well known that if $k\not\in\mathbb{Z}$, the CS term is not consistent as it is not invariant under certain large gauge transforms. The inconsistency should be cured once the low energy (IR) degrees of freedom are properly included, namely the full theory
\beq
\label{eq:CSUVIR}
\mathcal{S}_{IR}[A]+\int d^3x \frac{k}{4\pi}AdA\quad,
\eeq
should be fully gauge invariant. A familiar situation is when $k=1/2$, where the IR theory can be a gapless Dirac fermion. 
The Dirac fermion can be represented in the (Euclidean time) path integral formulation as
\beqa
Z_{IR}&=&\int D\bar{\psi}D\psi\exp{\left(\int d^3x\bar{\psi}i\slashed{D}_A\psi\right)} \nonumber \\ &=&|\det(\slashed{D}_A)|\exp{\left(\frac{i\pi}{2}\eta[A]-\int d^3x \frac{i}{8\pi}AdA \right)},
\label{eq: singleDirac2D}
\eeqa
where  $\eta[A]$ is the $\eta$-invariant. We refer to Ref.~\cite{Witten16} for a detailed review of the $\eta$-invariant. Here we only emphasize that the $\eta$-invariant is classically similar to the $k=1/2$ CS term in terms of equation of motion (and hence Hall conductivity), but is fully gauge-invariant unlike the $k=1/2$ CS term. 
The IR theory of the Dirac fermion thus fulfills two important requirements: it carries the opposite gauge non-invariance with the $k=1/2$ CS term, with a vanishing contribution to the net Hall conductance.

From a field theoretic point of view, it is somewhat arbitrary to separate the theory into a CS term and $\mathcal{S}_{IR}$. In fact the Dirac fermion can be defined in a gauge-invariant way with just the $\eta$-invariant~\cite{Witten16}, for example using the Pauli-Villars regulator. However, it will be useful to clearly separate the two contributions since it allows interesting generalizations that will be discussed later in this paper: the CS term can be interpreted as a ``UV'' contribution that comes from integrating out high-energy degrees of freedom and should therefore be analytic (but not necessarily fully consistent); Eq.~\eqref{eq: singleDirac2D} is interpreted as an ``IR'' contribution. The IR contribution does not have to be analytic since it comes from gapless degrees of freedom (in this case the $\eta$-invariant is not analytic), and should restore gauge invariance while keeping the UV response (in this case Hall conductance) unchanged. 

The above story is similar to quantum anomalies, in the sense that the IR theory should be nontrivial and match certain gauge non-invariance condition. However it is also different from the standard quantum anomalies, since the gauge non-invariance is imposed by fine-tuning the Hall conductance to a fixed fractional value. This ``unquantized quantum anomaly'' is therefore not protected like the standard anomalies, in the sense that a general perturbation can in principle change the Hall conductance. However we can adopt a rule of game in which the Hall conductance is fixed, which is justified if it is measured directly from the experiments or numerics. Then gauge invariance will impose strong constraints on the IR phases in a way similar to the standard quantum anomalies. In particular, the constraints can be applied to strongly correlated system --- in this case it leads to the familiar result that a system with a fractional quantum Hall conductance, even with strong interactions, must necessarily be long-range entangled.

Our ``unquantized anomaly'' can be viewed as a type of gauge non-invariant counter terms. The most familiar example of such counter term in condensed matter physics is perhaps the diamagnetic term in metals: $(n/2m) |{\bf A}|^2$, where $n$ is the electron density, $m$ is the fermion mass and ${\bf A}$ is the electromagnetic vector potential. The analogue of the fractional Hall conductance discussed above would be the optical conductivity $\sigma(\omega)\sim1/i\omega$ due to this counter term. This term is obviously gauge non-invariant and in the case of metals it demands a nontrivial Fermi surface to restore gauge invariance. The difference between the diamagnetic term and the fractional CS term is that the former is non-invariant under general (small and large) gauge transforms while the latter is non-invariant only under large gauge transforms, and is therefore more ``topological". In the rest of this work we will focus on such ``topological'' counter terms.

The rest of this paper is devoted to generalizing the above story to a variety of topological semimetals. The theory of such topological semimetals can be written, in a similar fashion to Eq.~\eqref{eq:CSUVIR}, as
\beq
S_{IR}[\psi,\mathcal{A}]+S_{UV}[\mathcal{A}]\quad,
\eeq
where $S_{IR}[\psi,\mathcal{A}]$ represents the low energy degrees of freedom such as the gapless fermions, $S_{UV}$ is a topological response term, which is a generalization of the Hall conductance, and $\mathcal{A}$ represents the probe gauge fields of the relevant symmetries, such as $U(1)$ and lattice symmetries. The inconsistency of $S_{UV}$ implies that the topological semimetal phase must remain long-range entangled even with strong interactions, as long as the topological response from $S_{UV}$ is fixed.

The relation between the standard and unquantized anomalies can be made more precise through the notion of \textit{emergent anomalies}.  The IR theory often enjoys a symmetry larger than that of the microscopic system. This emergent IR symmetry, which we denote as $G_{IR}$, can have some nontrivial t'Hooft anomalies, in the sense that if we formally couple the IR theory to a probe $G_{IR}$ gauge field, the theory is only sensible when viewed as the boundary of a bulk (denoted as $M$), with a nontrivial bulk response action $i\int_{M}w[G_{IR}]$ where $w[G_{IR}]$ is the corresponding topological term. We then re-insist that the true microscopic symmetry $G_{UV}$ is smaller, and is implemented in the IR theory as a subset of $G_{IR}$ through a map (a homomorphism)
\beq
\varphi: G_{UV}\to G_{IR}\quad,
\eeq
which gives a map (a pullback) $\varphi^*$ from the IR anomaly $w[G_{IR}]$ to the anomaly of the UV symmetry
\beq
w[G_{UV}]=\varphi^*w[G_{IR}]\quad.
\eeq
The unquantized anomaly we discuss here corresponds to the situation where the above $w[G_{UV}]$ is a total derivative as a bulk term: $w[G_{UV}]=d\Omega[G_{UV}]$, so that the UV anomaly is trivial at the cohomology level. However, it still reduces to a nontrivial counter term on the boundary $\Omega[G_{UV}]$. $\Omega$ is the analogue of the unquantized CS term $AdA$, where the corresponding bulk term is just the theta term $w=dAdA$.

Before entering the detailed discussions, we shall first review the notion of gauge fields for lattice symmetries.

\subsection{Review of lattice symmetry gauge fields}
\label{subsec:LSGF}

For an ordinary on-site discrete symmetry $G$ (such as the Ising $\mathbb{Z}_2$), what the probe gauge field $\mathcal{A}$ measures is essentially the twisted boundary conditions around each space-time $1$-cycle $C$. Specifically, a nontrivial Wilson loop $\int_{C}\mathcal{A}=g\in G$ means that adiabatically travelling along $C$ for a full cycle is equivalent to acting on the system by $g$. Here $C$ could also be a small cycle around a gauge defect (a vortex), in which case $\int_{C}\mathcal{A}$ measures the flux trapped in the defect. If we view vortices (with nontrivial gauge flux) as defects in the continuum space-time, the discrete gauge field becomes locally flat in the continuum $d\mathcal{A}=0$. Mathematically this means that $\mathcal{A}\in H^1(M,G)$, the first cohomology group of $G$, where $M$ is the space-time manifold.

\begin{figure}[t]
    \centering
    \vspace{0cm}
        \includegraphics[width=0.8\columnwidth]{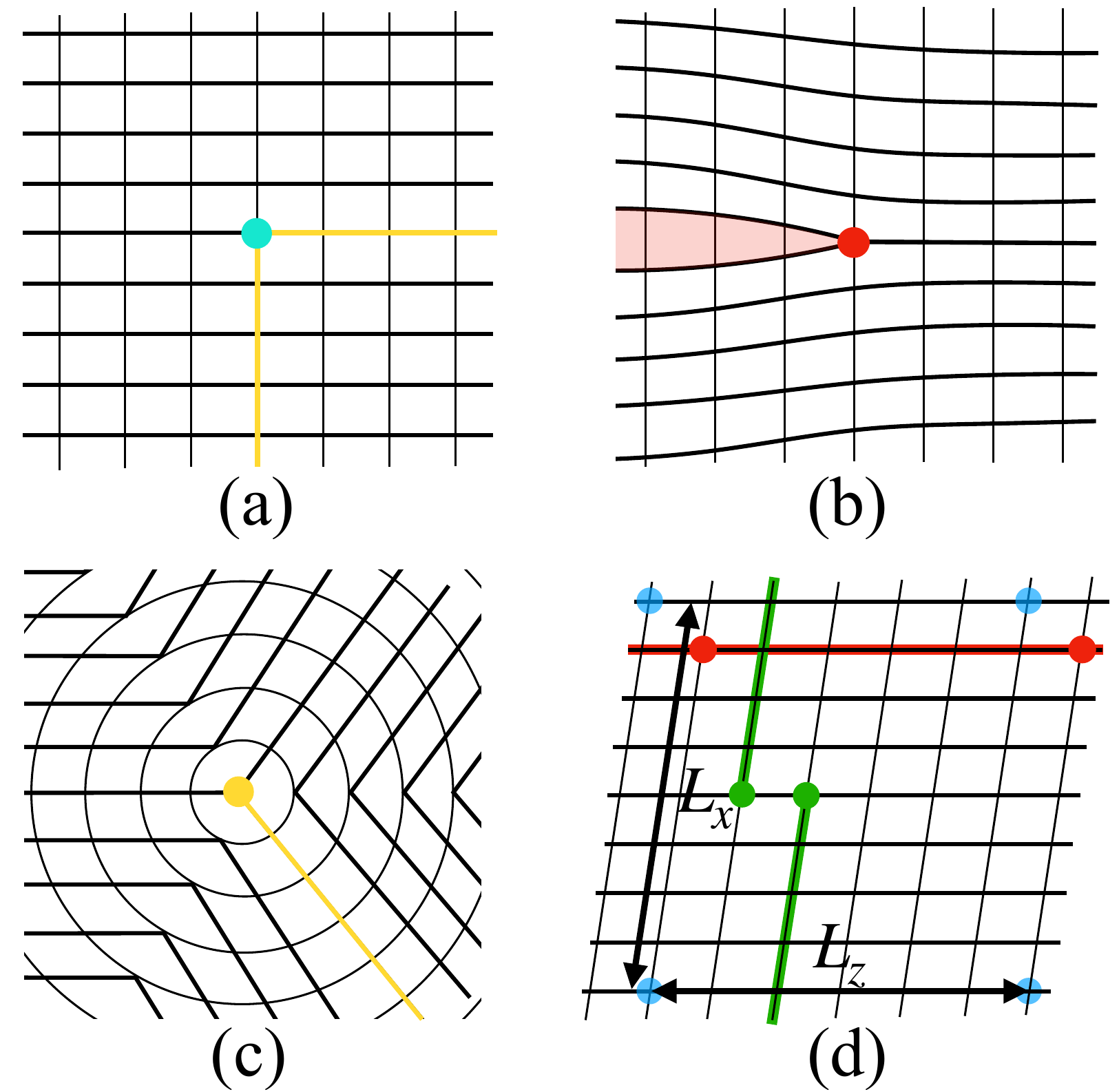}
    \caption{\label{fig:2ddefects} (Color online) Spatial symmetry point defects in 2D. (a) A defect-free 2D lattice with highlighted (blue) point possessing both translational and $\pi/2$ rotational symmetry. (b) A translational symmetry defect, known as a dislocation, is obtained by inserting an extra (red) half-plane and represented by the red dot. A Wilson loop around the defect gives $\int_C\mathcal{X}_i=1$. (c)  Gluing together the yellow lines in (a) produces a $\pi/2$ rotational defect known as a disclination. (d) Here we depict a periodic 2D lattice in the $xz$-plane with linear size in the $z$-direction $\int_{z=0}^{L_z}z=L_z$ (red line) and a shear strain in $z$ given by $\int_{x=0}^{L_x} z=1$ (green line). The four blue dots are equivalent to each other due to the periodic boundary conditions.}
\end{figure}

The above definition can be generalized to lattice symmetries~\cite{ThorngrenElse}. We start with lattice translation symmetries in $d$ space dimensions, where the symmetry forms the group $\mathbb{Z}^{\otimes d}$. For each translation symmetry $T_{x_i}$ in the $i$'th direction ($1\leq i\leq d$), we introduce a $\mathbb{Z}$-gauge field $\mathcal{X}_i$. Just like the Wilson loops in ordinary gauge theories, the integer $\int_{C}\mathcal{X}_i$ measures the number of $\hat{x}_i$-translations one has to go through to travel across the $1$-cycle $C$. To be more concrete consider a path integral description, with dynamical degrees of freedom $\psi$ (bosonic or fermionic) defined in continuous time $t\in[0,T)$ and on discrete lattice sites $s$ in space:
\beq
e^{-iS_{eff}[A,x_i]}=\int D[\psi(s,t)]\exp{\left(-i\sum_s\int dt \mathcal{L}_s[\psi,A]\right)},
\eeq
where we have used locality and translation symmetries to write the Lagrangian as a sum of local terms of identical form, $\mathcal{L}_s[\psi,A]$, which involves only fields near site $s$. We take periodic boundary conditions in space and time (so $M$ is a torus). The translation gauge fields enter the partition function by specifying exactly how the periodic boundary conditions are taken: 
\beqa
\label{eq:translationTBC}
\psi(s,t)&=&\psi\left(s+\hat{x}_j\int_i\mathcal{X}_j,t\right); \nonumber \\
\psi(s,t)&=&\psi\left(s+\hat{x}_j\int_t\mathcal{X}_j,t+T\right).
\eeqa
We now explain these equations in more detail. The Wilson loop of $\mathcal{X}_i$ in the $\hat{x}_i$ direction gives the lattice size $\int_i\mathcal{X}_i=L_i$. For $j\neq i$ the number $\int_i\mathcal{X}_j$ measures how much the slice of the lattice at $x_i=L_i$ is displaced along the $\hat{x}_j$ direction before it is identified with the slice at $x_i=0$. Similarly the time component $\int_t\mathcal{X}_i$ measures the displacement of the entire lattice at $t=T$ before identified with $t=0$. In other words, while the ``longitudinal" parts of the translation gauge fields measure the lattice size, the ``transverse" parts measure the quantized shear strains of the lattice in both space and time. We can also consider a $(d-2)$ dimensional defect in space, around which $\int \mathcal{X}_i=n\neq0$: this is simply a lattice dislocation with Burgers vector $\vec{B}=n\hat{x}_i$. This relation can also be written as $\int_S d\mathcal{X}_i=N$ where $N$ is the total charge of dislocations penetrating through the $2$-surface $S$. This is illustrated in Fig.~\ref{fig:2ddefects}. 

In the above discussion the lattice is viewed as a set of discrete points, which could exist without referring to any microscopic continuum geometry. However it is often convenient to embed the lattice into a continuous space, so that each site $s$ can be assigned a continuous coordinate $\vec{u}(s)$. Following usual practice in elasticity, the lattice coordinate can be treated as a field in the continuous space-time by assigning to each point $\vec{r}$ the value of $\vec{u}(s(\vec{r}))$, where $s(\vec{r})$ is the lattice point closest to $\vec{r}$. In this case the translation gauge field $\mathcal{X}_i$ can be interpreted as the elasticity tetrad $\mathcal{X}_i=\vec{\nabla}u_i$ \cite{DZYALOSHINSKII198067}, since the tetrad also satisfies Eq.~\eqref{eq:translationTBC}. Different continuum embeddings of the lattice lead to different tetrad representations of the $\mathcal{X}_i$ gauge fields, but the Wilson loops $\int_C\mathcal{X}_i$ do not depend on the details of the embedding. In this work we focus on universal properties, such as the Wilson loops, that do not depend on how the lattice is embedded into a continuum space. For example, we do not discuss the couplings between local elastic deformations (such as phonons or local strains) and the electrons~\cite{Franz16,Grushin16}. This allows us to treat $\mathcal{X}_i$ purely as a $\mathbb{Z}$-valued gauge field, and view the tetrad representation as a ``gauge choice'' of the translation gauge fields.

We note that the concept of translation gauge fields and elasticity tetrad have been used in recent literature in various contexts, including three dimensional integer quantum Hall effect~\cite{Tetrads1,Tetrads2, Tetrads}, Weyl semimetals~\cite{torsion_huang,Wang20}, electric polarizations ~\cite{song2019electric,Nissinen20} and crystalline symmetry-enriched topological orders~\cite{Manjunath_2020}.

We can similarly introduce probe gauge fields for lattice rotation symmetries. For example for a lattice $C_n$ rotations ($n=2,3,4,6$), we can introduce a $\mathbb{Z}_n$ gauge field $c$. A defect around which $\int c=m\neq0$ (mod $n$) is simply a lattice disclination. This lattice rotation gauge field has been used recently to characterize certain crystalline topological phases \cite{Liu19}.

\section{Chiral anomaly in (1+1)d lattice systems}
\label{sec:1+1d}

\subsection{$U(1)\times \mathbb{Z}$ chiral anomaly}
\label{sec:Hall}

We begin by reviewing the well known chiral (or filling) anomaly in (1+1)-dimensional lattice systems, with the aim to generalize these concepts to more complex situations. 
This $(1+1)$d chiral anomaly may be viewed as the fundamental anomaly, from which the higher dimensional anomalies in topological semimetals, that we will be concerned with 
in this paper, follow. 

\begin{figure}[t]
    \centering
    \vspace{0.2cm}
        \includegraphics[width=\columnwidth]{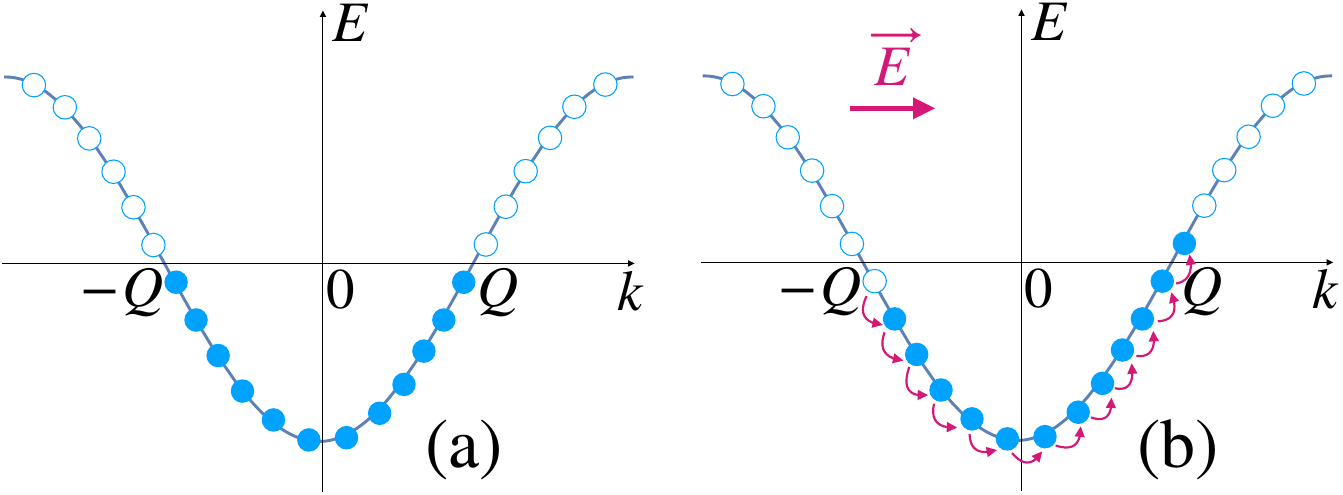}
    \caption{\label{fig:1ddispersion} (Color online) Illustration of the $(1+1)$d chiral anomaly: (a) A one band fermionic dispersion at fractional filling $\nu=2Q/2\pi$. (b) Once we thread a flux $\int dz A_z=2\pi$, which gives rise to an electric field $\vec{E}$, all filled states gain a unit of momentum resulting in a change of $2\pi\nu$ in chiral charge (i.e. crystal momentum).}
\end{figure}

Let us consider a $(1+1)$d ring of length $L_z$ with a single spinless fermionic band at a fractional filling $\nu$.  This system possesses the discrete lattice translational symmetry
$\mathbb{Z}$ and the $U(1)$ charge conservation symmetry. 
The band dispersion is shown in Fig.~\ref{fig:1ddispersion}(a). 
Luttinger~\cite{Luttinger60,Oshikawa00} or, more generally, Lieb-Schultz-Mattis~\cite{LSM,Oshikawa1999,Hastings04}), theorems tell us that this system is necessarily gapless in the presence of the $U(1)$ and translational symmetries. 
Alternatively, we may view this gaplessness as being mandated by a $U(1)\times\mathbb{Z}$ chiral anomaly that requires the existence of low-energy gapless modes in order to maintain gauge invariance~\cite{Furuya15,Cheng2015,Cho2017,Metlitski2017,Jian2017}.

This $U(1)\times\mathbb{Z}$ anomaly is a lattice descendant of the continuum $U(1)\times U(1)_a$ $(1+1)$d chiral anomaly, where $U(1)_a$ corresponds to the chiral (or axial) symmetry group. In the presence of a $U(1)$ flux there follows a non-conservation of the chiral charge, despite the presence of the chiral symmetry (hence \textit{anomaly}). In the lattice system this continuous chiral symmetry is replaced by the $\mathbb{Z}$ translational symmetry and as such leads to a non-conservation of the $\mathbb{Z}$ translational charge (which is simply the crystal momentum) when treated as an on-site symmetry. 
This is most easily demonstrated if we adiabatically thread a magnetic flux $\Phi=\oint dz A_z$ through the center of the ring, which causes a change in $\Phi=0$ momentum given by
\begin{align}
\partial_t P_{tot}&=\nu \int dz\, \partial_tA_z\quad,
\end{align}
where $\nu=2Q/2\pi$ with $2Q$ being the momentum separation between the chiral modes. 
For notational simplicity we will set the lattice constant $a$ to unity henceforth. Upon an insertion of $\Phi=2\pi$ we have
\begin{align}
\frac{\Delta P_{tot}}{2\pi}&= \nu\,\, (\mathrm{mod}\,\mathbb{Z})\quad,
\label{eq:CCnoncons}
\end{align}
where $\mathrm{mod}\,\mathbb{Z}$ arises due to the crystal momentum being only defined modulo a reciprocal lattice vector. This demonstrates non-conservation of the chiral charge, i.e. momentum, when $\nu\neq0\,(\mathrm{mod}\,\mathbb{Z})$ and is illustrated in Fig.~\ref{fig:1ddispersion}.

Another facet of the chiral anomaly has to do with the overall $U(1)$ charge of the ground state, which is given by
\beq
Q_{U(1)}=\sum_{k_z=-Q}^Q1=\nu L_z\quad,
\label{eq:chiralmagnetic11}
\eeq
where $L_z$ is the length of the system, which means that the charge per unit cell is $\nu$. Any noninteger value of $\nu$ is incompatible with a trivial gapped insulator state. An intuitive way to see this is to notice that the total $U(1)$ charge in Eq.~\eqref{eq:chiralmagnetic11} is
not properly quantized for some $L_z$ if $\nu\notin\mathbb{Z}$. This means that some additional charge $\delta Q\sim O(1)$ has to supplement Eq.~\eqref{eq:chiralmagnetic11} to make the charge quantized, no matter how large $L_z$ becomes. Furthermore, this $\sim O(1)$ additional charge cannot come from a trivially gapped state, since it is  non-analytic with respect to $1/L_z$ -- for example, an acceptable example will be $\delta Q=\lfloor \nu L_z\rfloor-\nu L_z$, which is badly non-analytic in $1/L_z$. The correction $\delta Q$ then has to come from some long-range entanglement -- in our case the gapless fermions. Although the relation Eq.~\eqref{eq:chiralmagnetic11} is rather trivial, its generalizations, which will be discussed extensively later, are not.

Both of the above manifestations of the $(1+1)$d chiral anomaly may be compactly expressed in terms of the following action, involving the translation (chiral) and the 
electromagnetic gauge fields
\beq
S=\nu\int  z\wedge A = \nu \int dtdz \,\epsilon^{\mu\nu}z_\mu A_\nu\quad,
\label{eq:chiralanomaly11}
\eeq
where $A$ is the usual $U(1)$ gauge field, $z\in H^1(\mathcal{M},\mathbb{Z})$ is the $\mathbb{Z}$ translational gauge field, and $\epsilon^{\mu\nu}$ is the Levi-Civita tensor in $(1+1)$d. 
All discrete gauge fields should be locally flat in the continuum limit since discrete fluxes lead to singular points in space. For the $z$ gauge field this means that $\int_{\mathcal{C}_2}dz=0$ when integrated over any closed $2$-cycle $\mathcal{C}_2$. Around a $1$-cycle $\mathcal{C}$, $\oint_{\mathcal{C}} z$ counts the number of $z$ translations traversed by the cycle, which is generally zero unless the loop is non-contractable, i.e. encloses omitted points in space. For example if we choose the cycle $\mathcal{C}_z$ to be along the length of the system then we obtain $\oint_{\mathcal{C}_z}z=L_z$. In general $\int_i z_j$, where $i\neq j$ measures the number of $z$ lattice slice displacements that are traversed along a cycle from $x_i=0$ and $x=L_{x_i}$. Let us now show how this action term reproduces the previously discussed physics of the chiral anomaly.

Recall that the minimal coupling between the current $j^\mu$ and gauge field $A_\mu$ is given by $-j^\mu A_\mu$. This means that when we vary Eq.~\eqref{eq:chiralanomaly11} with respect to the time component of the gauge field, $A_t$, we arrive at the expression for the total $U(1)$ charge as shown in Eq.~\eqref{eq:chiralmagnetic11}.
When we vary with respect to the time component of the translation gauge field, $z_t$, we get the total ground state momentum
\beq
P_{tot}(\Phi)=-\oint_{\mathcal{C}_z}\frac{\delta S}{\delta z_t}
=-\nu\Phi\quad.
\eeq
The momentum difference between the $\Phi=0$ and $2\pi$ ground states is then given by
\beq
\frac{P_{tot}(2\pi)-P_{tot}(0)}{2\pi}=-\nu\,\, (\mathrm{mod}\,\mathbb{Z}) \quad.
\label{eq:groundstateP}
\eeq
We note that the apparent sign difference between Eq.~\eqref{eq:groundstateP} and \eqref{eq:CCnoncons} is in fact consistent. Eq.~\eqref{eq:groundstateP} describes the change of \textit{ground state momentum} in the presence of a $2\pi$-flux, while Eq.~\eqref{eq:CCnoncons} is the momentum carried by the low energy (particle-hole) excitation induced by an adiabatic $2\pi$-flux insertion. The two should sum to zero since the process of adiabatic flux-insertion commutes with lattice translation and should not induce an actual momentum change. In our language (discussed in Sec.~\ref{sec:prelim}) Eq.~\eqref{eq:groundstateP} can be interpreted as a ``UV'' response since it is fixed by the lattice-scale information (the charge filling), and Eq.~\eqref{eq:CCnoncons} can be interpreted as the ``IR'' contribution since it originates from gapless particle-hole excitations of the IR theory.

The incompatibility of the anomaly action Eq.~\eqref{eq:chiralanomaly11} with a trivial insulator ground state may be clearly seen by examining how the action transforms 
under large gauge transformations. For example, if $A_t$ is taken to be spatially constant and wind by $2 \pi n$ around the temporal cycle, the corresponding contribution 
to the action is given by
\beq
\label{eq:largegauge}
S = - 2 \pi \nu n L_z\quad,
\eeq
which is generally nontrivial. This contradicts the fact that such a $2 \pi n$ winding of $A_t$ may be generated by a gauge transformation, i.e. Eq.~\eqref{eq:chiralanomaly11} 
is not gauge invariant. This means that there must exist gapless modes, which compensate for this gauge non-invariance.

We now comment on the relation between the ``unquantized anomaly'' Eq.~\eqref{eq:chiralanomaly11} and the standard t'Hooft anomaly. Although physically the only exact symmetry we impose here is $U(1)\times \mathbb{Z}$, at low energy the emergent Dirac fermion possesses an emergent $U(1)_c\times U(1)_a$ symmetry (the charge and axial charge conservation). If this $U(1)_c\times U(1)_a$ symmetry is exact and on-site, the system can only be defined on the edge of a $(2+1)d$ ``quantum spin Hall insulator'' bulk. This means that when coupled to a $U(1)_c$ gauge field $A$ and a $U(1)_a$ gauge field $B$, the Dirac fermion must be defined together with a mutual Chern-Simons (CS) term in one higher dimension:
 \beq
 \frac{1}{\pi}\int_{X_3} d^3xBdA\quad,
 \eeq
where the Dirac fermion lives on the boundary $\partial X_3$. For our example, $A$ is the electromagnetic gauge field. The axial charge is nothing but the crystal momentum, so we should set $B=-k_Fz$ -- here we temporarily treat the gauge field $z$ as continuous (defined in $\mathbb{R}$ instead of $\mathbb{Z}$), and recall that the charges under $z$ are nothing but the momenta of the fermion modes $\pm k_F$. Using the Luttinger theorem $2k_F=2\pi\nu$ and the fact that $dz=0$ for $z\in H^1(X_3,\mathbb{Z})$ (the $z$ gauge field is discrete at the end of the day), the total anomaly becomes
\beq
-\nu\int_{X_3}zdA=\nu\int_{\partial X_3}z\wedge A\quad,
\eeq
which is just Eq.~\eqref{eq:chiralanomaly11}. Importantly, the $BdA$ anomaly becomes trivial as a bulk term, but on the boundary it produces a nontrivial counter term which forces the IR theory to be nontrivial.

Now we will demonstrate how this basic $(1+1)$d chiral anomaly may be generalized to more complex situations, involving other crystalline symmetries, such as rotations.

\subsection{$\mathbb{Z}\times\mathbb{Z}_2$ anomaly}
\label{sec:rotationalcharge}
Discrete symmetries, such as discrete lattice rotations and translations, do not admit local charge densities (in contrast to $U(1)$). Nevertheless the charges of these discrete symmetries are globally defined and much of the discussion from the previous example can be generalized accordingly. We now discuss the simplest example with lattice $\mathbb{Z}$ translation in $\hat{z}$ direction and an on-site $\mathbb{Z}_2$ symmetry. For later use we interpret this $\mathbb{Z}_2$ as a $C_2$ rotation around the $\hat{z}$ axis.

\begin{figure}[t]
    \centering
    \vspace{0cm}
        \includegraphics[width=0.6\columnwidth]{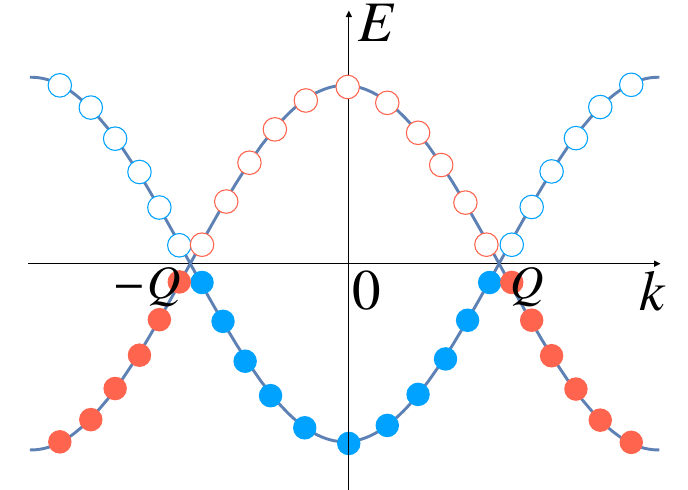}
    \caption{\label{fig:1ddispersioncz} (Color online) Band dispersion corresponding to the Hamiltonian in Eq.~\eqref{eq:11drotham}. The filled red and blue states correspond to different $C_2$ eigenvalue states. The sum of the individual filled charges gives the total $C_2$ charge which may be non-trivial, leading to a $\mathbb{Z}\times\mathbb{Z}_2$ chiral anomaly.}
\end{figure}

Consider a $(1+1)$d spinful fermionic square lattice model with translational symmetry in $z$, and $C_2$ symmetry, described by the following Hamiltonian
\begin{align}
H&=\frac{1}{2}\sum_{i}\left(c^\dag_{ i}\sigma^z c_{i+1 }-m\,c^\dag_{i}\sigma^z c_{i}+h.c.\right)\quad,\nonumber\\&=\sum_{k}\left(\cos k-m\right)c_{k}^\dag\sigma^zc_{k}\quad,
\label{eq:11drotham}
\end{align}
where $\sigma$ corresponds to the spin-degree of freedom, and we have two zero energy nodes at momentum $k=\pm Q$, with $Q=\cos^{-1}\left(m\right)$. The dispersion is shown in Fig.~\ref{fig:1ddispersioncz}. The gaplessness of the band dispersion is protected by the combination of $C_2= \sigma^z$ symmetry and translational symmetry. 
The total $C_2$ charge $Q_{C_2}$, defined through the $C_2$ eigenvalue $e^{iQ_{C_2}}$ for the many-body ground state, is determined by the filling fraction $\nu=2Q/2\pi$ of the band with $C_2$ eigenvalue $-1$:
\beq
Q_{C_2}=\pi\nu L_z+O(1)\quad,
\label{eq:C2charge1d}
\eeq
where $\nu=2Q/2\pi$ describes the separation between the band-touching nodes. 
A trivial symmetric gapped ground state with $C_2$ may only be a
product state in $C_2$ charges of either $0$, or $\pi$ at every sites. Thus the total $C_2$ charge of a trivial symmetric system can correspond to $0$ or $\pi L_z$ respectively. We see that a trivial state is unable to capture the total charge of the system for a fixed general nodal separation of $2Q/2 \pi\notin \mathbb{Z}$. One can also heuristically understand the ``anomaly'' of Eq.~\eqref{eq:C2charge1d} in a similar way as the previous example: the $C_2$ charge in Eq.~\eqref{eq:C2charge1d} is not properly quantized for certain $L_z$ when $\nu\notin\mathbb{Z}$, so an additional $\delta Q_{C_2}\sim O(1)$ has to be added. In general $\delta Q_{C_2}$ will be non-analytic in $1/L_z$ and therefore should come from some nontrivial IR modes, like the gapless fermions in our example. 

Parallel to the $U(1)\times\mathbb{Z}$ anomaly case, this inability to form a trivial state at certain nodal separations is the result of a $\mathbb{Z}\times \mathbb{Z}_2$ chiral filling anomaly associated with the $z$ translation and $C_2$ symmetries.
Analogous to the $U(1)\times\mathbb{Z}$ anomaly, the effect of the $C_2$ charge is encoded in the following topological term
\beq
S=\pi\nu\int z\wedge c\quad,
\label{zcanomaly}
\eeq
where $c\in H^1(\mathcal{M},\mathbb{Z}_2)$ is the $C_2$ gauge field which is the on-site spin rotation gauge field.\footnote{Strictly speaking we should be using the cup product $\cup$ for discrete gauge fields. But for most purposes in this paper it suffices to consider the standard wedge product $\wedge$.}
The general trivial states correspond to $\nu\in \mathbb{Z}$.

By construction, when we vary the action in Eq.~\eqref{zcanomaly} with respect to the time component of the $C_2$ gauge field, $c_t$, we arrive at the $Q_{C_2}$ in agreement with Eq.~\eqref{eq:C2charge1d}.
In addition, varying with respect to $z_t$, we obtain
\beq
\label{eq:zcanomalymomentum}
P_{tot}(\Phi_c)=\oint_{\mathcal{C}_z}\frac{\delta S}{\delta z_t}=-\pi\nu\Phi_c\quad,
\eeq
where $\Phi_c=\oint_{\mathcal{C}_z} c_z$ is the $C_2$ flux. This means that a nontrivial $C_2$ flux (a periodic boundary condition twisted by $C_2$) induces a nontrivial momentum. What appears inconsistent is that even a trivial flux $\Phi_c=2$ also induces a nontrivial momentum: 
\beq
\frac{P_{tot}(2)-P_{tot}(0)}{2\pi}=-\nu\,\, (\mathrm{mod}\,\mathbb{Z}) \quad.
\eeq
The resolution is that the trivial $\Phi_c=2$ flux induces a gapless excitation with momentum $-2\pi\nu=-2Q$ -- this is nothing but a particle-hole excitation near the Fermi points. This provides another physical reason for the existence of nontrivial IR modes.

The formal way to express the above anomaly, just as in the case of the $U(1) \times \mathbb{Z}$ anomaly, is to notice that the topological term in Eq.~\eqref{zcanomaly} is not gauge invariant under large gauge transformations $c\to c+2\alpha$ where $\alpha$ is an integer $1$-form, 
which mandates gapless low-energy modes to restore gauge invariance. 
As we will see later, this anomaly has a $(3+1)$d extension, which corresponds to rotation symmetry protected (type-I) Dirac semimetals. 

The $\Z_2\times\Z$ anomaly Eq.~\eqref{zcanomaly} does have a subtle aspect not present in the $U(1)\times\Z$ case. We are always free to redefine the $\Z_2$ gauge field $c\to (2n+1)c$ for any $n\in\Z$. Therefore different values of $\nu$ should give identical response (or anomaly) if they differ by a factor of $2n+1$, which means the following equivalence relation
\beq
\label{eq:equiv}
\nu\sim (2n+1)\nu, \quad n\in\Z.
\eeq
This equivalence relation can be understood physically from either Eq.~\eqref{eq:C2charge1d} or Eq.~\eqref{eq:zcanomalymomentum}. In Eq.~\eqref{eq:C2charge1d} we can multiply the total charge by any odd integer without changing the physical meaning, since the total charge $Q_{C_2}/\pi$ is only defined in $\Z_2$. Likewise in Eq.~\eqref{eq:zcanomalymomentum}, we can multiply the flux $\Phi_c$ by any odd integer without changing the physics since $\Phi_c$ is defined in $\Z_2$.  

An immediate consequence of the equivalence relation Eq.~\eqref{eq:equiv} is that the anomaly is in fact trivial if
\beq
\label{eq:trivialnu}
\nu=\frac{n}{2m+1}\sim n, \quad n,m\in\Z.
\eeq
These \textit{exceptional values} form a measure zero but dense subset within the interval $[0,1]$. We can also demonstrate the triviality of these values of $\nu$ more explicitly by constructing a trivial phase starting from the metallic state: we can first gap out all the fermions by introducing a charge-density-wave (CDW) order that breaks the $\Z$ translation symmetry but keeps the $\Z_2$. For the values in Eq.~\eqref{eq:trivialnu} the CDW order parameter lives in $\Z_{2m+1}$. We can recover the $\Z$ translation symmetry by proliferating (condensing) domain walls of the $\Z_{2m+1}$ CDW order parameter. A standard calculation (see Appendix~\ref{app:exceptional}) shows that the domain wall (denoted as $\sigma$) formally carries $\Z_2$ charge $\pi/(2m+1)$, namely $\Z_2:\sigma\to e^{i\pi/(2m+1)}\sigma$. We can combine this with a $\Z_{2m+1}$ gauge transform for the domain wall $\sigma\to e^{i2m\pi/(2m+1)}\sigma$ and realize that the domain wall in fact carries only an integer charge\footnote{Mathematically this is the familiar statement that $\Z_2$ symmetry cannot be fractionalized on a $\Z_{2m+1}$ gauge-charged particle, or $H^2(\Z_2,\Z_{2m+1})=0$.} of $\pi$ under $Z_2$. We can therefore neutralize this $\Z_2$ charge by attaching to the domain wall a local operator that is odd under $\Z_2$. This way we obtain a domain wall operator $\sigma'$ which can now be condensed without breaking any other symmetry, and the resulting state is fully symmetric and gapped. Notice that if we had $\nu=1/2^n$ instead, the domain wall will carry $\Z_2$ charge $\pi/2^n$ which is now truly fractional, i.e. it cannot be eliminated through gauge transforms and attaching local operators. In this case the domain walls cannot be proliferated without breaking $\Z_2$ symmetry, and a gapped symmetric phase is impossible.

The anomaly and exceptional points can also be understood using the notion of emergent anomalies\cite{Metlitski2017}, which we briefly discuss in Appendix~\ref{app:exceptional}.

Another consequence of the equivalence relation Eq.~\eqref{eq:equiv} is that the magnitude of $\nu$ is no longer meaningful. In fact any value of $\nu$ can be made arbitrarily close to either $0$ or $1$ by appropriately multiplying some factor $(2n+1)/(2m+1)$. For rational values of $\nu$, we can uniquely write $\nu=2^np/q\sim 2^n$ with $p,q$ odd and $n\in\Z$. Curiously, we can write this relation more compactly as
\beq
\nu\sim |1/\nu|_2, \quad {\rm if}\quad \nu\in\mathbb{Q},
\eeq
where $|...|_p$ denotes the $p$-adic magnitude.

Finally we notice that the $\mathbb{Z}_2\times\mathbb{Z}$ anomaly is unambiguisly defined only if the $U(1)\times \mathbb{Z}$ filling anomaly is trivial (namely the system has integer charge filling). Otherwise we can make a large $U(1)$ gauge transform $A\to A+2\pi Nc$ ($N\in\mathbb{Z}$) in the filling anomaly Eq.~\eqref{eq:chiralanomaly11}, which results in a shift of the coefficient of the $\mathbb{Z}_2\times\mathbb{Z}$ anomaly.

\subsection{$\mathbb{Z}\times\mathbb{Z}$ anomaly}
\label{sec:ZZanomaly}

The final $(1+1)$d example we want to discuss is the most nontrivial and involves four chiral modes, two right-handed and two left-handed, protected only by the translational 
symmetry. The $U(1)$ and $C_2$ symmetries can still be there but are not important for the following discussion. This situation will be relevant to the time-reversal invariant Weyl semimetal, which will be discussed in Sec.~\ref{sec:TRinvariantWSM}.

\begin{figure}[t]
    \centering
    \vspace{0cm}
        \includegraphics[width=0.6\columnwidth]{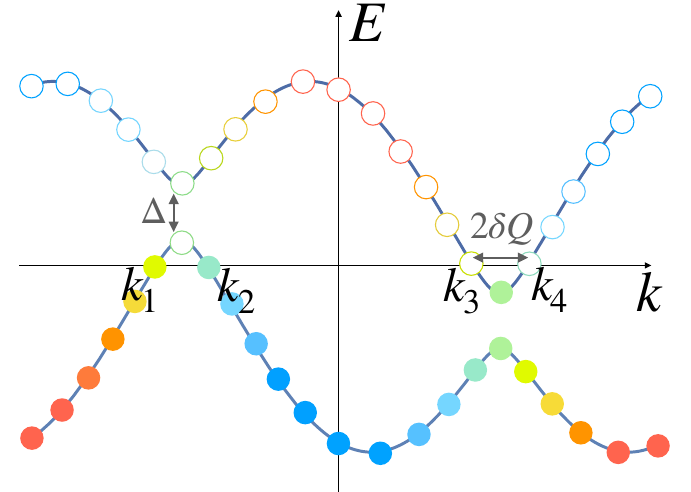}
    \caption{\label{fig:1ddispersionzz} (Color online) The blue (red) band in Fig.~\ref{fig:1ddispersioncz} is shifted to the right (left) by $\delta Q$. $C_2$ symmetry is not necessary for the protection of the states anymore as indicated by the colour hybridisation. Instead, the filled states now possess a non-trivial total momentum, leading to a $\mathbb{Z}\times\mathbb{Z}$ chiral anomaly.}
\end{figure}

We consider a modified model of the previous $(1+1)$d spinful fermionic system Eq.~\eqref{eq:11drotham}, in which both the $C_2$ rotation and inversion symmetries are broken, 
so that only the translational symmetry remains
\beqa
\label{eq:ZZHamiltonian}
H&=&\sum_{\langle i,j\rangle}\left[\cos \delta Q c^\dg_{ i}\sigma^z c^\pdg_j -m\,c^\dg_{i}\sigma^z c^\pdg_{i} + \Delta c^\dg_i \sigma^x c^\pdg_i \right. \nonumber \\
&-&\left. \frac{i}{2} \sin \delta Q  (c^\dg_{i}c^\pdg_{j} - c^\dg_j c^\pdg_i) \right] \nonumber \\
&=&\sum_{k}c_{k}^\dg\left[\left(\cos\delta Q \cos k - m\right)\sigma^z + \Delta \sigma^x - \sin \delta Q \sin k \right]c^\pdg_k. \nonumber \\
\eeqa
We now have have positive chirality gapless nodes at $\pm k_+$ and negative chirality gapless nodes at $\pm k_-$, with $k_\pm=Q\pm\delta Q$ (taking the $C_2$ symmetry breaking 
parameter $\Delta$ to be negligible for simplicity). This is essentially a shifted version of the two overlayed bands studied in the previous section, shown in Fig.~\ref{fig:1ddispersioncz}, where each band is moved by $\delta Q$ in opposite directions as shown in Fig.~\ref{fig:1ddispersionzz}. The low energy modes in this model are solely protected by translational symmetry in the $z$ direction, which suggests the existence of an anomaly relating to the translational gauge field. 

A key feature of this shifted dispersion is the existence of total ground state $z$-component of the crystal momentum given by
\beq
P_{z}=\sum_{k_z = k_-}^{k_+}k_z-\sum_{k_z=-k_+}^{-k_-}k_z=\pi\lambda L_z+O(1)\quad,
\label{eq:totalP}
\eeq
where $\lambda=2Q\delta Q/\pi^2$ (defined mod $2$). This relation is similar to Eq.~\eqref{eq:chiralmagnetic11} and \eqref{eq:C2charge1d}, with the symmetry charge being the total momentum. Similar to the previous examples, the $O(1)$ piece in Eq.~\eqref{eq:totalP} is required for proper momentum quantization: $2\pi L_zP_z\in\mathbb{Z}$. For $\lambda=1$ the $O(1)$ piece can simply be $\delta P_z=\pi$. For fractional $\lambda\notin\mathbb{Z}$ one can show that the $O(1)$ piece has to take some nontrivial form, in particular it can not be analytic in $1/L_z$. This means that for $\lambda\in[0,1)$ some nontrivial IR modes (like gapless fermions) are needed. Equivalently, a short-range entangled ground state must have $\lambda$ equal to either $0$ or $1$ -- the latter can be realized by a product state with an odd number of fermions in each unit cell.

We comment on a subtlety with the above discussion. One may worry that the coefficient $\lambda$ in Eq.~\eqref{eq:totalP} is not well defined since $P_z$ is only defined mod $2\pi$. In particular, we may arbitrarily shift $\lambda\rightarrow\lambda+\eta$ by simply adding a factor $2\pi \left\lfloor\eta L\right\rfloor$ to the sequence. One way to view this issue is to regard different choices of $\lambda$ as being different choices of Brillouin zones over which we do our ground state summation in Eq.~\eqref{eq:totalP}. Thus our job is to fix such a summation convention that allows us to uniquely determine  $\lambda$.

To specify how $\lambda$ is determined we demand the following condition on the momentum choice:
\begin{align}
    \frac{P(L)}{2\pi}\leq \frac{P(L+1)}{2\pi}\leq \frac{P(L)}{2\pi}+1\,, && \quad P(2)=\pi\,,
    \label{eq:condition}
\end{align}
which corresponds to determining ground state momentum via purely counting momenta within $0$ and $2\pi$. These two demands uniquely determine the sequence of $P_s(L)$ as $L\rightarrow\infty$ and thus a unique $\lambda$. 
Using such a sequence of ground state momenta $P_s(L)$, we may then determine $\lambda$ via the procedure
\begin{align}
    \lambda\equiv\lim_{L\rightarrow \infty}\frac{P_s(L)}{\pi L}\quad.
    \label{eq:lambdadetermination}
\end{align}

To further elucidate the process, let us first consider the trivial case of a fully-filled band with one electron per unit cell. If we choose periodic boundary conditions we may obtain an exact expression for the ground state momentum $P(L)=\pi L-\pi$, where $-\pi$ is the $\mathcal{O}\left(1\right)$ term in Eq.~\eqref{eq:totalP}. Such a $P(L)$ automatically satisfies conditions in Eq.~\eqref{eq:condition}, and via procedure \eqref{eq:lambdadetermination} yields $\lambda=1$ as expected. It is important to note that in this case the $\mathcal{O}\left(1\right)$ term is of an analytic nature in $1/L$ which expresses the trivial nature of the system and allows for an insulator to reproduce such a behaviour. However for non-trivial systems the $\mathcal{O}\left(1/L\right)$ piece comes about due to the difference between a summation of filled state momenta at $L$, and the momentum integral in the thermodynamic limit. This pattern is highly configuration and length dependent due the difference between the gapless mode momenta defined in the thermodynamic limit and the corresponding well-defined filled momentum at $L$. This sort of difference behaves non-analytically in $1/L$ and can only be feasibly reproduced by a gapless or long-range entangled state.

To further support the nontriviality of the $\mathbb{Z}\times\mathbb{Z}$ anomaly, in Appendix~\ref{sec:LLAC} we perform an explicit stability analysis using Luttinger liquid theory. 
Now we continue by discussing the action that expresses the presence of a total gauge-invariant ground state momentum.

\subsubsection{Anomaly term}
\label{sec:ZZanomalyterm}
Expressing the $\mathbb{Z} \times \mathbb{Z}$ anomaly as a topological term is a little more subtle than in the previous two examples. If the spacetime forms a simple torus $T^2$, i.e. simple periodic boundary condition in both time and space, then the following term can reproduce the response in Eq.~\eqref{eq:totalP}:
\beq
\label{eq:zzresponse}
S=\pi\lambda\int z_tz_z\quad,
\eeq
where $z_t$ and $z_z$ are the time and space components of the $z$ gauge field, respectively. This action does not appear to be topological, nor does it show any gauge non-invariance since $z\in H^1(\mathcal{M},\mathbb{Z})$ does not have any large gauge transform. The resolution is to notice that translation symmetry is in fact different from an on-site $\mathbb{Z}$ symmetry in one important aspect: on a finite-size system with length $L_z$, the many-body total momentum\footnote{This is true when there is no symmetry flux (twisted boundary condition) from other symmetries, which is justified here since we are focusing on just the translation symmetry.}  is quantized $P_z\in \frac{2\pi}{L_z}\mathbb{Z}$. Since $L_z=\oint_{\mathcal{C}_z} z_z$, the momentum quantization imposes a somewhat unconventional large gauge symmetry on a spacetime torus $T^d$:
\begin{align}
z\to & z+\left(\oint_{\mathcal{C}_z} z_z\right)\alpha, \nonumber \\
\oint_{\mathcal{C}_z}\alpha &=0,\quad\oint_{\mathcal{C}_{i\neq z}}\alpha\in\mathbb{Z}.
\end{align}
This gauge symmetry can be understood from the definition of the translation gauge field reviewed in Sec.~\ref{subsec:LSGF}: the ``transverse'' part of the gauge field $\oint_{\mathcal{C}_{i\neq z}}\alpha$ measures the displacement in $\hat{z}$ direction as one moves around the $\mathcal{C}_i$ cycle, and this displacement is defined only modulo $L_z$.
This large gauge symmetry also allows us to have more nontrivial gauge field configurations (bundles) with nontrivial $\int_{\mathcal{C}_2}dz$ over some 2-cycles $\mathcal{C}_2$. Specifically, we allow $\int_{\mathcal{C}_2}dz\in L_z\mathbb{Z}$ for $\mathcal{C}_2$ not including any $\mathcal{C}_z$ cycle, and  $\int_{\mathcal{C}_2}dz=0$ if $\mathcal{C}_2$ includes a $\mathcal{C}_z$ cycle. This immediately implies that $\int dz=0$ in $(1+1)d$, which is relevant for our discussion here, and $\int dzdz=0$ in $(3+1)d$ which will be relevant for our later discussion in Sec.~\ref{sec:TRinvariantWSM}.

We can now see why the response term Eq.~\eqref{eq:zzresponse} is not gauge invariant: under $z_t\to L_z\alpha$, the term changes by $\delta S=N\pi\lambda L_z^2$ for some integer $N$. For $\lambda=1$ this change can be made trivial by further supplementing a counter term $\pi\int z_t$ which changes by $N\pi L_z$ (recall that $L_z(L_z+1)$ is always even). For fractional $\lambda$ there is no such counter term, so the gauge non-invariance is intrinsic and nontrivial IR modes are required to cancel this gauge non-invariant. 

Similar to the two previous examples, we can also formally write the response Eq.~\eqref{eq:zzresponse} as the boundary descendent of a bulk term
\beq
\pi\lambda\int_{X_3}zdz=\pi\lambda\int_{X_3}z_z(\partial_uz_t-\partial_tz_u)=\pi\lambda\int_{\partial X_3}z_tz_z,
\eeq
where $u$ is the direction perpendicular to the boundary. The equality follows from the requirements on $\int_{\mathcal{C}_2} dz$ discussed above. The topological nature of the response is more manifest in this $\int zdz$ form -- the price we pay is that it is defined in one higher dimension.

We can further illustrate how the $\int zdz$ term appears through the standard chiral anomaly analysis. A chiral fermion, coupled to a continuous ($U(1)$ or $\mathbb{R}$) gauge field $A$, can only be defined on the boundary of a bulk with nontrivial Hall conductance:
\beq
\label{ChernSimons}
S = \pm \frac{1}{4 \pi} \int A \wedge d A\quad,
\eeq
where the sign is determined by the chirality of the fermion. In our current example, we should replace $A\to k_zz$, where $k_z$ is the crystal momentum of the chiral fermion. For the theory defined in Eq.~\eqref{eq:ZZHamiltonian}, the full anomaly is
\beqa
\label{zzanomaly}
S&=&\frac{1}{4 \pi} \int \left[ (k_+ z) \wedge d (k_+ z)+(- k_+ z) \wedge d (- k_+ z) \right. \nonumber \\
&-& \left. (k_- z) \wedge d (k_- z)-( - k_- z) \wedge d ( - k_- z) \right] \nonumber \\
&=& \pi\lambda \int z \wedge d z\quad,
\eeqa
where we have used $\pi\lambda\equiv 2Q\delta Q/\pi=(k_+^2-k_-^2)/2\pi$.

Similar to the $\mathbb{Z}_2\times\mathbb{Z}$ anomaly, the $\mathbb{Z}\times\mathbb{Z}$ anomaly is unambiguisly defined only if the $U(1)\times \mathbb{Z}$ filling anomaly is trivial (i.e. integer charge filling). Otherwise we can make a large $U(1)$ gauge transform $A\to A+2\pi Nz$ ($N\in\mathbb{Z}$) in the filling anomaly Eq.~\eqref{eq:chiralanomaly11} to shift the coefficient of the $\mathbb{Z}\times\mathbb{Z}$ anomaly.

\section{Three-dimensional topological semimetals}
\label{sec:2}
We will now apply the results of Section~\ref{sec:1+1d} to $(3+1)$d semimetal systems. 
The connection between $(1+1)$d and $(3+1)$d anomalies is well-known: in an externally applied magnetic field, a $(3+1)$d system, 
exhibiting a chiral anomaly, possesses special lowest Landau levels (LLL), which exhibit the corresponding $(1+1)$d chiral anomaly.

\subsection{TR-broken Weyl semimetal}
\label{sec:2a}
Let us start with the system that may be viewed as the ``hydrogen atom" of topological semimetals, namely the simplest magnetic Weyl semimetal with a pair of nodes, located on the $z$-axis in momentum space at $k_z = \pm Q$. The gaplessness of such a system is protected purely by $U(1)$ charge symmetry and translational symmetry in the $z$ direction, and is well-known for possessing a chiral anomaly of the form~\cite{Zyuzin12-1,Burkov_ARCMP}
\begin{align}
S&=-\frac{1}{2}\frac{\nu}{2\pi}\int z\wedge A\wedge dA\quad,\nonumber\\
&=-\frac{1}{2}\frac{\nu}{2\pi}\int dt\,d^3 r\, z_{\mu}\epsilon^{\mu\nu\lambda\eta}A_{\nu}\partial_\lambda A_\eta \quad,
\label{eq:WSMaction}
\end{align}
with $\nu=\frac{2Q}{2\pi}$. 
Eq.~\ref{eq:WSMaction} encodes the standard topological responses, which have been discussed extensively before, i.e. the anomalous Hall conductivity $\sigma_{xy}=\frac{\nu}{2\pi}$, and the chiral magnetic effect (CME). 
In addition, it also encodes  more subtle responses, which directly probe translational symmetry defects and involve nontrivial charges on magnetic flux loops linked with crystal 
dislocations. It is this type of phenomena, generalizable to other kinds of topological semimetals, which do not possess any obvious electromagnetic topological responses, 
that will be the main focus of this paper. 

When a Weyl semimetal with a pair of nodes, separated by $2Q$ along the $z$-axis, is placed in an external magnetic field along the same direction, 
it develops a spin-polarized LLL, which disperses along the direction of the field and crosses the Fermi energy at the locations of the Weyl points, exactly as 
shown in  Fig.~\ref{fig:1ddispersion}(a). 
Thus the LLL of the magnetic Weyl semimetal in an external magnetic field maps directly onto a $(1+1)$d metal with the $U(1)\times\mathbb{Z}$ chiral anomaly, described 
in Sec.~\ref{sec:Hall}. Taking into account the LLL orbital degeneracy $N_{LLL}=B L_xL_y/2\pi $, where $L_{x,y}$ are the sample sizes in the $x$ and $y$-directions, the derivative of the Luttinger volume of this $(1+1)$d metal with respect to the magnetic field gives the Hall conductivity of the Weyl semimetal. 
This is encoded in the topological response action in the $(t,z)$ plane
\beq
\label{eq:WSM_LLL}
S=\nu N_{LLL}\int z\wedge A\quad,
\eeq
which also follows directly from Eq.~\eqref{eq:WSMaction}. We may also invert this argument and say that Eq.~\eqref{eq:WSMaction} follows from Eq.~\eqref{eq:WSM_LLL}. This is the logic we will use to find topological response terms for other semimetals, for which no purely electromagnetic responses, like the Hall effect, exist. 
Namely we will identify topological response terms by mapping the LLL of these semimetals to one of the $(1+1)$d systems, discussed in Section~\ref{sec:1+1d}.

The incompatibility of the anomalous responses, described above, with a trivial gapped insulator may also be seen explicitly if one attempts to construct such an 
insulator starting from a gapped superconductor and disordering the phase of the superconducting order parameter by proliferating vortices. 
As discussed in Refs.~\cite{Wang20,Thakurathi20}, a gapped superconducting state can only be obtained in this case using FFLO-type pairing (for weak pairing), where electrons on each side of the two Weyl nodes are paired and the pairs thus carry momentum $\pm 2Q$~\cite{Meng12,Moore12,Bednik15,YiLi18}. 
This generally breaks crystal translational symmetry, except when $2 Q = \pi$, i.e. half a reciprocal lattice vector.

Consider, say, a right-handed Weyl fermion with a singlet superconducting pairing. 
The pairing Hamiltonian is given by
\beq
\label{eq:1}
H = \sum_\bk c^\dg_\bk \bsigma \cdot \bk c^\pdg_\bk + \Delta \sum_\bk (c^\dg_{\bk \upa} c^\dg_{-\bk \da} + c^\pdg_{- \bk \da} c^\pdg_{\bk \upa})\quad.
\eeq
Here $\bsigma$ are Pauli matrices corresponding to the degree of freedom describing the two bands that touch at the Weyl point 
and the momentum $\bk$ is measured from the location of the Weyl point $Q \hat z$. We have also set the Fermi velocity of the Weyl fermion to unity. 
Introducing the Nambu spinor notation $\psi_\bk = (c^\pdg_{\bk \upa}, c^\pdg_{\bk \da}, c^\dg_{- \bk \da}, - c^\dg_{- \bk \upa})$, this may be rewritten as
\beq
\label{eq:2}
H = \frac{1}{2} \sum_\bk \psi^\dg_\bk (\tau^z \bsigma \cdot \bk + \Delta \tau^x) \psi^\pdg_\bk\quad,
\eeq
where the Pauli matrices $\tau^a$ act on the particle and hole components of the Nambu spinor. 
Apart from a factor of $1/2$ in front of the sum over momenta, correcting for the doubling of degrees of freedom in the Nambu representation, Eq.~\eqref{eq:2} has the form of the Hamiltonian of a free Dirac fermion of mass $\Delta$.

Now consider a straight-line vortex of positive unit vorticity along the $z$-axis, i.e. we take the superconducting order parameter to have the 
following form in cylindrical coordinates $\Delta(\br) = |\Delta(r)| e^{i \theta}$, where $r = \sqrt{x^2 + y^2}$ and $\theta = \arctan(y/x)$. 
The momentum-space Hamiltonian in Eq.~\eqref{eq:2} is replaced by the following Bogoliubov-de Gennes (BdG) Hamiltonian
\beqa
\label{eq:3}
\cH&=&- i \tau^z (\sigma^x \partial_x + \sigma^y \partial_y) + \tau^z \sigma^z k_z \nonumber \\
&+&|\Delta(r)| (\cos \theta \tau^x - \sin \theta \tau^y)\quad.
\eeqa
This is a classic problem first considered in a different context by Callan and Harvey~\cite{CallanHarvey}. 
Taking first $k_z = 0$ and looking for a $\theta$-independent localized solution of the BdG equation
\beq
\label{eq:4}
\cH \Psi = 0\quad,
\eeq
one obtains, ignoring normalization factor
\beqa
\label{eq:5}
\Psi_R(\br) = \left(
\begin{array}{c}
1\\0\\0\\-i 
\end{array}
\right) 
e^{- \int_0^r d r' |\Delta(r')|}\quad,
\eeqa
where the subscript $R$ refers to the right-handed Weyl fermion of Eq.~\eqref{eq:1}. 
It is easy to see that $\Psi_R(\br)$ is also an eigenstate of the BdG Hamiltonian~\eqref{eq:3} at a nonzero $k_z$ with eigenvalue $k_z$, 
i.e. it describes a right-moving mode, localized in the vortex core. 
Repeating the same calculation for the left-handed Weyl fermion we have
\beqa
\label{eq:6}
\Psi_L(\br) = \left(
\begin{array}{c}
1\\0\\0\\i 
\end{array}
\right) 
e^{- \int_0^r d r' |\Delta(r')|}\quad,
\eeqa
which is an eigenstate of the corresponding BdG Hamiltonian with eigenvalue $- k_z$, i.e. it corresponds to a left-moving mode, localized in the vortex core.

These chiral Majorana modes in the vortex cores prevent a gapped insulating state if the superconducting phase coherence is destroyed by phase fluctuations. 
This conclusion holds for any odd number of pairs of Weyl nodes, i.e. for any magnetic Weyl semimetal.
To obtain an insulating state, we need to condense vortices with higher vorticity, the different possible states are discussed in detail in Refs.~\cite{Wang20,Thakurathi20,Sehayek20}. The vortex condensation method's inability to obtain a trivial symmetric gapped insulator with non-trivial Hall conductivity perfectly reflects the anomalous nature of the magnetic Weyl semimetal. 
In the following subsection we will generalize this analysis to other topological semimetals, which do not have any nontrivial electromagnetic responses, and thus the answer to the question of what exactly is topological about them is far less obvious.

\subsection{Type-I Dirac semimetal}
\label{sec:DSM}
We will now attempt to generalize the analysis in the previous subsection to the case of type-I Dirac semimetals. 
Type-I Dirac semimetals are TR and parity invariant which guarantees doubly degenerate bands. Their band dispersions feature a pair of Dirac nodes, located at time-reversed momenta on an axis of rotation, where each node consists of a pair of overlapping negative and positive chirality Weyl nodes. Their gaplessness is protected by a combination of rotational, translational, and $U(1)$ charge conservation symmetries. 
In a close analogy to magnetic Weyl semimetal, one may think of a type-I Dirac semimetal as an intermediate phase between an ordinary insulator and a weak TR-invariant topological 
insulator, where the direction of the weak index coincides with the rotation axis. 

For concreteness we will consider a specific realization of a type-I Dirac semimetal with four-fold rotational symmetry, described by the following Hamiltonian in momentum space~\cite{Fang12}
\beqa
\label{eq:12}
&&\cH(\bk) = \sin k_x\, \Gamma_1 + \sin k_y \, \Gamma_2 + m(\bk) \Gamma_3  \\
&+&\gamma_1 (\cos k_x - \cos k_y) \sin k_z\, \Gamma_4 + \gamma_2 \sin k_x \sin k_y \sin k_z\, \Gamma_5. \nonumber 
\eeqa
Here 
\beq
\label{eq:13}
m(\bk) = m_0 - b_{xy} (2 - \cos k_x - \cos k_y) - b_z (1 - \cos k_z), 
\eeq   
and the $4 \times 4$ matrices $\Gamma_a$, satisfying Clifford 
algebra $\{\Gamma_a, \Gamma_b\} = 2 \delta_{ab}$, are defined as 
\beq
\label{eq:14}
\Gamma_1 = \sigma^x s^z, \,\, \Gamma_2 = - \sigma^y, \,\, \Gamma_3 = \sigma^z,\,\, 
\Gamma_4 = \sigma^x s^x,\,\, \Gamma_5 = \sigma^x s^y, 
\eeq 
where the $2 \times 2$ Pauli matrices $\boldsymbol{\sigma}$ and $\bs$ refer to the orbital parity and the spin degrees of freedom 
respectively.
The parameters in the function $m(\bk)$ are assumed to be chosen in such a way that band inversion occurs at the 
$\Gamma$-point in the BZ, producing two Dirac points along the $k_x = k_y = 0$ axis, whose location is 
given by 
\beq
\label{eq:15}
k_z = \pm Q = \pm \arccos\left(1 - m_0/b_z\right)\quad. 
\eeq 
As mentioned above, this Dirac semimetal state may be regarded as an intermediate phase between an ordinary insulator when 
$m_0 < 0$ and a weak topological insulator with the weak indices $(0,0,1)$ when $m_0 > 2 b_z$ and $b_{xy} > b_z$. 

As we have chosen to define Eq.~\eqref{eq:12} on a cubic lattice $\cH(\bk)$ possesses a C$_4$ rotational symmetry about the $z$-axis, with the rotation operator given by 
$R_4 = e^{i\pi/4} e^{- \frac{i \pi}{4} (2 - \sigma^z) s^z}$, where 
the factor $e^{i\pi/4}$ was insterted for later convenience. 
This rotational symmetry is what protects the Dirac points, since it prohibits mass terms $\sin k_z \Gamma_{4,5}$, which are odd under 
rotation
$R_4^\dg \Gamma_{4,5} R_4^\pdg = - \Gamma_{4,5}$.
Additionally $z$ translation symmetry prevents prohibits a charge density wave from opening a gap, and $U(1)$ charge conservation symmetry prevents superconductivity.

Topologically nontrivial properties of type-I Dirac semimetals are related to the presence of gapless chiral fermions at the Dirac nodes, just as in the simpler magnetic Weyl 
semimetal case discussed above. Analogously to the magnetic Weyl case, it is then useful to place the system in an external magnetic field along the $z$-direction, 
which accomplishes an effective dimensional reduction to a $(1+1)$d problem.

\begin{figure}[t]
    \centering
    \vspace{0cm}
        \includegraphics[width=0.5\columnwidth]{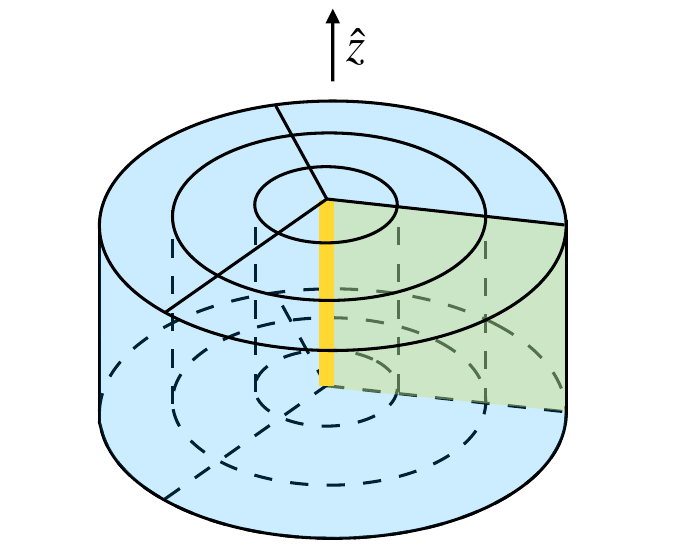}
    \caption{\label{fig:disclination} (Color online) A $C_4$ disclination with the Frank vector along $\hat z$ is depicted by gluing two lattice faces, rotated by $\pi/2$ with respect to each other, together (shown as yellow surface) to create a defect line shown as a bold yellow line. Close to this defect the lattice appears distorted, however far away the regular square lattice shape is retained.}
\end{figure}

Since type-I Dirac semimetals contain two pairs of Weyl nodes at low energies, there is a pair of LLL with the following dispersion (see Appendix~\ref{app:LLL} for detailed 
derivation)
\beq
E^\pm_{LLL}(k_z)=\pm m(0,0,k_z)\quad,
\eeq
with $E^+_{LLL}$ having a $C_4$ charge of $0$, and $E^-_{LLL}$ having a $C_4$ charge of $\pi$.
One can see that the form of the LLLs is identical to the $1D$ band dispersions discussed in Section~\ref{sec:rotationalcharge} (see Fig.~\ref{fig:1ddispersioncz}), except that here we consider $C_4$ symmetry and there exists an additional Landau level degeneracy of $N_{LLL}=B L_xL_y/2\pi$ per band. 
The logic we used in Section~\ref{sec:rotationalcharge} may then be applied directly, after taking into account the LLL degeneracy. 
This implies the following $(1+1)$d topological term (with $\nu\equiv 2Q/2\pi$)
\beq
\label{eq:LLLc4}
S=\pi\nu N_{LLL}\int z\wedge c_4\quad,
\eeq
where $c_4\in H^1(\mathcal{M},\mathbb{Z}_4)$ is the $C_4$ gauge field which around a 1-cycle counts number of $C_4$ disclinations that are traversed -- an example of such a disclination is seen in Fig.~\ref{fig:disclination}. 

Extending Eq.~\eqref{eq:LLLc4} back to $(3+1)$ dimensions we obtain the generalized chiral anomaly term, which characterizes type-I Dirac semimetals
\beq
\label{eq:Diracanomaly}
S=\frac{\nu}{2}\int z\wedge c_4\wedge dA\quad.
\eeq

Unlike the chiral anomaly term in magnetic Weyl semimetal, Eq.~\eqref{eq:Diracanomaly} cannot be interpreted as a purely electromagnetic response (like the Hall conductance). In the Landau level interpretation Eq.~\eqref{eq:LLLc4}, a nontrivial $C_4$ charge per length in the $\hat{z}$ direction is associated with a magnetic flux in the $xy$-plane. Alternatively, we can also consider a $C_4$ disclination (Fig.~\ref{fig:disclination}(a)) in the $\hat{z}$ direction ($\int_{\mathcal{C}_{xy}} dc_4=1$). The action Eq.~\eqref{eq:Diracanomaly} reduces to the $(1+1)d$ filling anomaly Eq.~\eqref{eq:chiralanomaly11} with charge density $\nu/2$. This fractional charge density induced in the disclination is also a characterization of the anomaly of the type-I Dirac semimetal. The anomalous nature is manifest if we consider a ``trivial'' four-fold disclination ($\int_{xy}dc_4=4$), which now carries a charge density $2\nu$ and is fractional if $\nu$ takes a nontrivial value. The fractional charge associated with disclinations in symmorphic Dirac semimetals may alternatively be approached from the viewpoint of hinge states in the non-interacting limit~\cite{PhysRevB.99.245151,Wieder_2020}. The anomalous response in Eq.~\eqref{eq:Diracanomaly} offers a generalisation of this feature to include both translation and rotational charge responses, valid in both the free-fermion and strongly-correlated regimes.

The response Eq.~\eqref{eq:Diracanomaly} is invariant under large gauge transformations $c_4\to c_4+4\alpha$ if $\nu\in\frac{1}{2}\Z$. Similar to the $\Z_2\times\Z$ anomaly discussed in Sec.~\ref{sec:rotationalcharge}, there is a set of equivalence relations $\nu\sim (4n+1)\nu$ for any $n\in\Z$. We therefore conclude that the response is anomalous unless $\nu$ belongs to the set of \textit{exceptional values}
\beq
\label{eq:exceptionalmomentum}
\nu_{ex}=\frac{n}{2(2m+1)}, \quad n,m\in\Z.
\eeq
We note that at the level of free fermion band structure, the theory appears to be nontrivial at these exceptional values, with Dirac nodes at $k_z=\pm \pi\nu_{ex}$. Our analysis indicates that with strong interactions, a type-I Dirac semimetal with $\nu_{ex}$ can form a symmetric short-range entangled state. The resulting insulating state has a gauge invariant response Eq.~\eqref{eq:Diracanomaly} with $\nu=\tilde{n}/2$, where $\tilde{n}=(-1)^{m}n$. For $n\neq 0$ (mod $4$) this is a nontrivial crystalline SPT state, and can be viewed as a stack (in $\hat{z}$ direction) of $(2+1)d$ insulators with charge $\tilde{n}$ sitting at the $C_4$ rotation centers. The topological response of such $(2+1)d$ insulators has been discussed in Refs.~\cite{Liu19,Song2018}.

As in the magnetic Weyl case, it is also useful to see how an attempt to construct a trivial insulator by gapping the Dirac nodes fails, barring the exceptional points Eq.~\eqref{eq:exceptionalmomentum}, explicitly via the vortex condensation method. 
For this purpose it is convenient to focus on low-energy states and expand Eq.~\eqref{eq:12} to linear order in the transverse momentum components $k_{x,y}$, which gives
\beq
\label{eq:hlinear1}
\cH(\bk) = k_x \sigma^x s^z - k_y \sigma^y \pm (k_z \pm Q) \sigma^z\quad, 
\eeq
where we have absorbed the Fermi velocity along the $z$-direction into the definition of $k_z$. 
An ordinary gapped BCS superconducting state is now possible by pairing right- and left-handed fermions separately. 
For the pair of right-handed fermions we have
\beq
\label{eq:BCS1}
H = \sum_\bk c^\dg_\bk \bsigma \cdot \bk c^\pdg_\bk + \Delta \sum_\bk (c^\dg_{\bk 1} i \sigma^y c^\dg_{-\bk 2} + h.c.)\quad, 
\eeq
where we have brought the right-handed node Hamiltonian to the form $\bsigma \cdot \bk$ by a unitary transformation, which also 
changes the rotation operator to $R_4=e^{\frac{i\pi}{4}}e^{-\frac{i\pi}{4}(\sigma^z-2s^z)}$; $k_z$ is measured from the corresponding node 
location and the $1,2$ index labels the two eigenvalues of $s^z$, which distinguish the two right-handed nodes in the pair. 
Introducing Nambu spinor $\psi^\pdg_\bk = (c^\pdg_{\bk 1 \upa}, c^\pdg_{\bk 1 \da}, c^\dg_{-\bk 2 \da}, -c^\dg_{-\bk 2 \upa})$ this becomes
\beq
\label{eq:BCS2}
H = \sum_\bk \psi^\dg_\bk (s^z \bsigma \cdot \bk + \Delta s^x) \psi^\pdg_\bk\quad,
\eeq
which represents two identical copies of Eq.~\eqref{eq:2}, describing a single superconducting Weyl fermion. 
 Thus in this case we obtain a pair of Majorana, or a single chiral right-moving Dirac mode in the vortex core. 
 Analogously, the left-handed pair of Weyl nodes produces a single left-moving Dirac mode. 
 From Eqs.~\eqref{eq:5} and \eqref{eq:6} both of these modes are linear combinations of $c_{\bk 1 \upa}$ and $c^\dg_{- \bk 2 \upa}$. 
 This means that they transform under $C_4$ rotation as
 \beq
 \label{eq:rotations}
 R_4:\, \psi^{R,L}_{k_z} \ra i \psi^{R,L}_{k_z}\quad,
 \eeq
 where $R_4=e^{\frac{i\pi}{4}(2-\sigma^z+s^z)}$ in the Nambu basis.
 This means that any pairing of the left- and right-handed modes of the type
\beq
\label{eq:1dpairing}
H = \sum_{k_z} \left[k_z \psi^\dg_{k_z} \tau^z \psi^\pdg_{k_z} + \frac{\Delta}{2} \left(\psi^\dg_{k_z} i \tau^y \psi^\dg_{-k_z} + h.c. \right) \right]\quad,
\eeq
where the eigenvalues of $\tau^z$ label the chirality of the 1D Weyl modes, which would gap them out, violates the $C_4$ symmetry.
We then have to consider non-perturbative ways to gap out the fermions. For this it is convenient to first introduce a CDW order
\beq
m\psi^{\dagger}_L\psi_R+h.c.\quad,
\eeq
which gaps out the fermions by breaking translation symmetry. We then ask if translation symmetry can be restored by condensing defects of $m$. A calculation similar to that in Sec.~\ref{sec:rotationalcharge} and Appendix~\ref{app:exceptional} shows that this is possible only if the momenta of the nodes $\pm Q=\pm \pi\nu$ belong to the exceptional values Eq.~\eqref{eq:exceptionalmomentum}.\footnote{There is an additional subtlety with the exceptional value $\nu=1/2$. A direct calculation shows that the $\Z_2$ domain wall of the CDW order parameter in the vortex core carries $C_4$ charge $\pi/4$, which appears to be fractional and therefore disallows vortex condensation. However, notice that a bulk Bogoliubov fermion (not the vortex zero modes) sees the vortex as a $\pi$-flux, so if a vortex sits on a $C_4$ axis the bulk fermion will carry $C_4$ angular momentum $(2n+1)\pi/4$ ($n\in\Z_4$). So the $\pi/4$ charge associated with the vortex CDW domain wall can be canceled by bring in a bulk fermion into the vortex. The domain wall can then be condensed to give a trivially gapped vortex, which can then be gapped and produce a symmetric insulator.}
Thus, in a $C_4$ symmetric state the fermion modes remain gapless and single vortex condensation is impossible unless the Dirac nodes sit at some exceptional momenta defined in Eq.~\eqref{eq:exceptionalmomentum}.

\subsection{TR-invariant Weyl semimetal}
\label{sec:TRinvariantWSM}
Finally, let us discuss the most nontrivial case, that of a TR-invariant Weyl semimetal. 
In a TR-invariant Weyl semimetal the nodes always occur in multiples of four since there are pairs of nodes of equal chirality, related to each other by TR.
We will first discuss an example in which all the nodes lie on the $z$-axis, which is closely related to the $(1+1)$d system, discussed in Section~\ref{sec:1+1d}. 
We will then generalize to the situation when the nodes are not on the same line. 
\subsubsection{Nodes separated in the $z$ direction}
The simplest model for a TR-invariant Weyl semimetal may be obtained from the model of a type-I Dirac semimetal Eq.~\eqref{eq:12} by adding a $C_4$ symmetry breaking 
perturbation $\gamma \sigma^x s^x$ and an inversion-breaking perturbation $g \sin k_z \sigma^z s^z$. 
The Weyl node locations $k_z = \pm Q \pm \delta Q$ are nontrivial solutions of the equation
\beq
\label{eq:Weylnodes}
g^2 \sin^2 k_z = \gamma^2 + m^2(k_z)\quad.
\eeq
When a magnetic field is applied along the $z$-axis, the resulting LLL structure is identical to the $(1+1)$d system, discussed in Section~\ref{sec:ZZanomaly}, and shown in Fig.~\ref{fig:1ddispersionzz}.

\begin{figure}[t]
    \centering
    \vspace{0cm}
        \includegraphics[width=0.5\columnwidth]{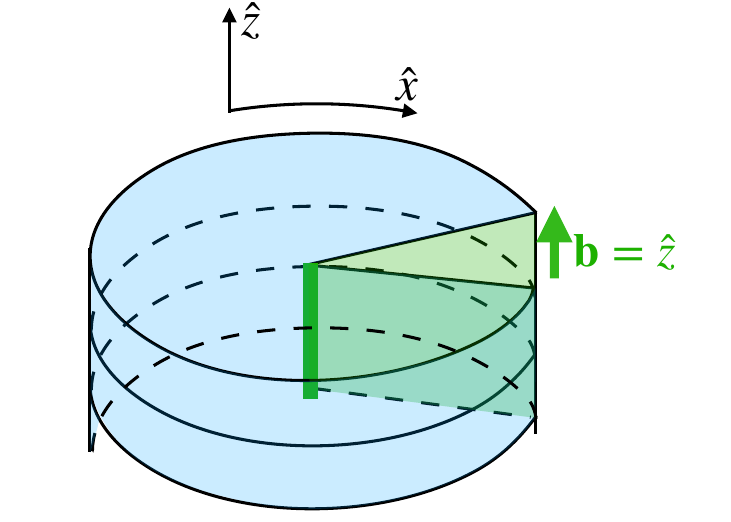}
    \caption{\label{fig:screwdislocations} (Color online) Cartoon of a screw dislocation, represented by a lattice shear strain along the green surface. The defect line is shown in bold green with a Burgers vector $\mathbf{b}=\hat{z}$.}
\end{figure}

Extending the $(1+1)$d anomaly action Eq.~\eqref{zzanomaly} to $(3+1)$ dimensions, we then obtain\footnote{Under a small gauge transform $A\to A+d\alpha$, the invariance of this action is guaranteed by the relation $\int dzdz=0$ discussed in Sec.~\ref{sec:ZZanomalyterm}.} 
\beq
\label{eq:TIWSMresponse}
S=-\frac{\lambda}{2}\int z\wedge dz\wedge A\quad. 
\eeq
where $\lambda=2Q\delta Q/\pi^2$. This action describes two distinct manifestations of the nontrivial topology of a TR-invariant Weyl semimetal. 
One is the nontrivial ground state momentum, which appears in the presence of an external magnetic field. 
This may be obtained by varying the action with respect to the time component of the translation gauge field $z_t$ 
\beq
P_z = \pi\lambda L_z N_{LLL}\quad,
\eeq
where $\mathcal{N}=\mathcal{C}_x\times \mathcal{C}_y\times \mathcal{C}_z$ with $\mathcal{C}_\alpha$ being the circumference cycle along the direction $\hat{\alpha}$, 
$\int_{\mathcal{C}_x\times\mathcal{C}_y} dA/2\pi= B L_xL_y/2\pi = N_{LLL}$, and we have taken $dz=0$ assuming a perfect crystal without dislocations. We see that the total momentum is exactly the expected momentum output as in Sec.~\ref{sec:ZZanomaly} with $N_{LLL}$ copies from the LLL degeneracy. (ii) The second physical phenomenon can be seen with a screw dislocation where $\int_{\mathcal{C}_{xy}}dz=b$, with the magnitude of Burgers vector ${\bf b} = \hat z$ (Fig.~\ref{fig:screwdislocations}). Varying with respect to $A_t$ gives a fractional $1D$ charge density on the screw dislocation
\beq
\rho=-\frac{\lambda}{2}b\quad.
\eeq

We comment on the role of time-reversal symmetry here. The response term Eq.~\eqref{eq:TIWSMresponse} does not require time-reversal. But if time-reversal is broken, a Hall conductance term Eq.~\eqref{eq:WSMaction} can be induced as the Weyl nodes can shift in asymmetric ways. A large gauge transform $A\to A+2\pi Nz$ ($N\in\mathbb{Z}$) will then shift the coefficient of the term Eq.~\eqref{eq:TIWSMresponse}. The anomaly is therefore sharply defined only in the absence of Hall conductance.

Analogously to the previous cases, topologically nontrivial nature of the TR-invariant Weyl semimetal also manifests in the impossibility of gapping out the Weyl nodes without either breaking translational symmetry or inducing topological order. 
The analysis here is essentially identical to the type-I Dirac semimetal case above and we will not show the details for this reason. 
One starts with a gapped BCS state, obtained by separately pairing right- and left-handed Weyl fermions at time-reversed momenta, described 
by Eq.~\eqref{eq:BCS1}. 
A $\pi$-flux vortex then binds right- or left-moving 1D Weyl fermion modes, which transform nontrivially under translations, since the modes exist 
at nonzero momenta, corresponding to the locations of the Weyl nodes. 
Since the left- and right-handed Weyl nodes are located at different momenta, not related to each other by any symmetry, pairing the corresponding 
left- and right-moving 1D Weyl fermion modes in the vortex core is impossible without breaking translational symmetry. 

\subsubsection{Nodes separated in the $xz$-plane}

Weyl nodes in a TR-invariant Weyl semimetal in general are not located on the same line, as in the special example considered above. It is, however, straightforward to extend the above analysis to a more general situation. 

For example, consider a system with four Weyl nodes, where the right-handed nodes are at momenta $\mathbf{k}=\pm(\delta Q,0,Q)$ and the left-handed ones are at $\mathbf{k}=\pm(-\delta Q,0,Q)$. In such a configuration, an applied magnetic field in the $x$ direction causes the LLL's to carry a total momentum of
\begin{align}
P_x=\pi\lambda L_z N_{LLL}\quad,
\end{align}
where $\lambda=2Q\delta Q/\pi^2$,
while a magnetic field in the $z$ direction has LLL's with total  momentum
\begin{align}
P_z=\pi\lambda L_x N_{LLL}\quad.
\end{align}
Such a response can be described by the action 
\beq
S=-\frac{\lambda}{2}\int \left(x\wedge dz+z\wedge dx\right)\wedge A\quad,
\eeq
where $x$ is the gauge field corresponding to $x$ translational symmetry. It is important to point out that this term is distinct from the electric polarization action, which has a superficially similar form $P\int x\wedge z\wedge dA$~\cite{song2019electric} since the momentum response of the Weyl semimetal is symmetric under $x\leftrightarrow z$, rather than the antisymmetric response from polarization. A polarization response is fully gauge invariant in the bulk and known to be realizable by a short-range entangled insulator and is thus not anomalous. In contrast, the TR-invariant Weyl semimetal has an anomalous response. Similarly to all previous cases of semimetals the anomalous nature is also exhibited by the fractional charge density, carried by a dislocation with Burgers vector $b\hat{z}$ or $b\hat{x}$. 

All other cases can be easily generalized from the two presented TR-invariant WSM examples. The most general TR-invariant WSM will have nodes shifted in all three dimensions and be described by a combination of different translational gauge field terms $\sum_{i,j}\int \gamma_{ij} x_i\wedge dx_j\wedge A$, where $(x_1,x_2,x_3)=(x,y,z)$. 
For a related recent work, discussing topological responses in TR-invariant WSM, see Ref.~\cite{dubinkin2021higher}.

\section{Discussion and conclusion}
\label{sec:conclusion}

As we have demonstrated in multiple examples above, the anomalous response of $(3+1)$d symmetry-protected semimetals may be reduced to $(1+1)$d chiral anomalies of their lowest Landau levels in an external magnetic field, involving the relevant protecting symmetries. We have shown that all these $(1+1)$d anomalies in turn stem from an unquantized filling anomaly, which essentially specifies the amount of symmetry charge present in the ground state. The relevant symmetry charges are the $U(1)$ charge, crystalline angular momentum and linear momentum for the cases of magnetic WSM, type-I DSM and TR-invariant WSM, respectively. These charges can be tuned to be trivial 
while preserving the symmetry of the system, hence leading to tunable quantum anomalies. However if we fix the anomaly prefactor, and thus the total charge, to be non-trivial, while demanding that the relevant symmetries remain unbroken, there must exist compensating IR behaviour such as gapless modes or topological order to maintain gauge invariance. It then becomes natural to view the gaplessness of $(3+1)$d semimetals as being topologically mandated by a non-trivial prefactor of a tunable anomaly, such as a non-integer (in appropriate units) Hall conductivity for magnetic WSM. In this work, we have extended this concept to cover other topological semimetal systems. The anomalies for the TR-invariant Weyl and type-I Dirac semimetals can also be characterized by fractional $U(1)$ charge densities, induced on crystalline symmetry defects, such as disclinations for type-I DSM and screw dislocations for TR-invariant WSM. To support the non-triviality of these semimetals at a non-perturbative level, we have also shown that a trivial gapped state with a fixed non-trivial anomaly prefactor is not achievable via vortex condensation.

These unquantized, tunable, anomalies naturally generalize the notion of fractional $U(1)$ charge density to other discrete symmetries (like the crystalline symmetries discussed in this work). In realistic systems, the $U(1)$ charge density (filling fraction) is naturally fixed by chemistry, while fixing the coefficients of other tunable anomalies, as one turns on electron-electron interactions, appears to be less natural and necessarily involves some fine-tuning. We can instead take a different viewpoint: given an interacting system in an unknown phase, we can in principle measure the coefficient of the tunable anomalies either experimentally or numerically, for example by measuring the symmetry charges induced in various defect configurations. We can then constrain the low energy theory (the phase) of this system from such measurement --- for example, if the system has a fractional $\sigma_{xy}$ in appropriate units, then it has to be either a metal, a magnetic Weyl semimetal, or a topologically ordered insulator.

Type-I Dirac semimetals with translational and $n$-fold rotational symmetry protected gapless states along the rotational axes were found to possess an unquantized $U(1)\times\mathbb{Z}\times\mathbb{Z}_{n}$ anomaly, except when the momentum separation between the nodes satisfies certain special conditions like Eq.~\eqref{eq:exceptionalmomentum} for $C_4$ rotation. Such \textit{exceptional points} form a measure zero but dense subset.

There are many additional questions that can be explored, such as whether there exist gapless $(3+1)$d systems that do not inherently involve $U(1)$ charge symmetry as opposed to the presented semimetal systems --- this would be relevant for the study of nodal superconductors. A well known example along this line is the magnetic Weyl semimetal with nodal seperation $2Q\neq N\pi$, which is nontrivial even without $U(1)$ symmetry due to the fractional thermal Hall conductivity. A generalization of these anomalies to systems protected by non-symmorphic symmetries (type-II Dirac semimetals) is also warranted. The fractional charge density, induced on various crystalline defects, predicted in this work can, at least in principle, be tested in experiments.

\begin{acknowledgments}
We acknowledge useful discussions with Ying Ran, Ashvin Vishwanath and Liujun Zou. LG was supported by the Natural Sciences and Engineering Research Council (NSERC) of Canada and by a Vanier Canada Graduate Scholarship. 
AAB was supported by Center for Advancement of Topological Semimetals, an Energy Frontier Research Center funded by the U.S. Department of Energy Office of Science, Office of Basic Energy Sciences, through the Ames Laboratory under
contract DE-AC02-07CH11358. 
Research at Perimeter Institute is supported in part by the Government of Canada through the Department of Innovation, Science and Economic Development and by the Province of Ontario through the Ministry of Economic Development, Job Creation and Trade.
\end{acknowledgments}

\begin{appendix}

\section{The $\Z_2\times\Z$ anomaly in $(1+1)d$: exceptional points and emergent anomalies}
\label{app:exceptional}

We analyze the low energy theory with four chiral fermions in the bosonized language. The Luttinger liquid consists of four compact bosons $e^{i \phi_I}$ ($I\in\{1,2,3,4\}$), with the Lagrangian
\begin{align}
\label{eq:LuttingerLagrangian}
\mathcal{L}=-\frac{1}{4\pi}\left[K_{IJ}\partial_t\phi_I\partial_z\phi_J+V_{IJ}\partial_z\phi_I\partial_z\phi_J\right]\quad,
\end{align}
where
\begin{align}
\label{eq:Kmatrix}
K=\begin{pmatrix}
1 & 0 & 0 & 0 \\
0 & 1 & 0 & 0 \\
0 & 0 & -1 & 0 \\
0 & 0 & 0 & -1
\end{pmatrix}\quad,
\end{align}
and $V=v_F\mathds{1}$ is the velocity matrix. The fermion creation operators are expressed as $\psi^\dag_I\sim \kappa^+ e^{i\phi_I}$, where $\kappa^+$ are the Klein factors, taking care of the anticommutation relations between different species of fermions. Upon the charge $U(1)$, translational $\Z$ and $\Z_2$ symmetry transformations, the boson phases change as
\begin{align}
\label{eq:LuttingerSymm}
U(1):&\quad\phi_I\rightarrow \phi_I + \theta\quad, \nonumber \\
\Z:&\quad\phi_I\rightarrow \phi_I +  k_I \quad,\nonumber \\
\Z_2:&\quad\phi_I\rightarrow \phi_I+\gamma_I\quad,
\end{align}
where $k_I=\pi\nu(1,-1,1,-1)$ and $\gamma_I=\pi(1,0,0,1)$.

We now gap out the fermions by breaking the translation symmetry, keeping $U(1)\times\Z_2$, through a CDW order parameter
\beq
\label{eq:CDWapp}
m[e^{i(\phi_1-\phi_4)}+e^{i(\phi_3-\phi_2)}]+h.c.\quad,
\eeq
Under translation $m\to e^{i2\pi\nu}m$. For $\nu\in\mathbb{Q}$, we write $\nu=p/q$ with coprime $p,q\in\Z$. Then $m$ takes value in $\mathbb{Z}_q$. A domain wall of $m$ is defined as a nonlocal operator, such that the CDW operator Eq.~\eqref{eq:CDWapp} rotates by a phase $e^{i2\pi/q}$ when commuted with the domain wall. It is not hard to see that the appropriate choice of the domain wall operator is
\beq
\sigma=\exp\left[i\frac{(\phi_1+\phi_4)-(\phi_2+\phi_3)}{2q}\right]\quad.
\eeq
This operator transforms trivially under $U(1)$ (this is expected since the $U(1)\times\Z$ filling anomaly vanishes in this case), but under $\Z_2$ it transforms as 
\beq
\Z_2: \quad \sigma\to e^{i\pi/q}\sigma\quad.
\eeq
For even $q$, the above transformation signals fractional (or projective) $\Z_2$ symmetry charge on $\sigma$. For odd $q$, we can choose a different gauge, for example by demanding that under $\Z_2$: $\phi_{1,4}\to \phi_{1,4}+q\pi$. In this gauge $\Z_2: \sigma\to-\sigma$, so $\sigma$ no longer carries fractional charge under $\Z_2$. 

We can also consider irrational values of $\nu$ (incommensurate CDW). In this case the CDW order parameter $m$ lives on $U(1)$. If we try to disorder $m$ and recover translation symmetry, we should proliferate vortices of $m$. The vortex operator of $m$ is simply
\beq
V=\exp\left[i\frac{(\phi_1+\phi_4)-(\phi_2+\phi_3)}{2}\right]\quad,
\eeq
which transforms nontrivially under $\Z_2: V\to-V$. Notice vortices are local operators and we are not free to attach other local operators to it. Therefore the nontrivial action of $\Z_2$ on the vortex operator signals an obstruction to having a symmetric gapped phase.

The above discussions can be rephrased in terms of emergent anomalies\cite{Metlitski2017}. First, as the $U(1)$ gauge field does not explicitly appear in the anomaly Eq.~\eqref{zcanomaly}, we can view the anomaly as coming entirely from the charge neutral sector of the system. But since charge neutral objects are all bosonic in the system, we will only need to consider bosonic anomalies. The t'Hooft anomaly in $(1+1)d$ corresponds to symmetry-protected topological (SPT) phases in $(2+1)d$, which are classified by group cohomology $H^3(G,U(1))$~\cite{Wen_SPT,Wen_SPT2}. 

For rational $\nu=p/q$, the translation symmetry acts on the low energy theory effectively as a $\Z_q$ symmetry (up to some $U(1)$ gauge transforms). Neglecting $U(1)$ from now on, the effective symmetry group of the low energy theory is $\Z_2\times\Z_q$. In $(2+1)d$ the mutual anomaly between $\Z_2$ and $\Z_q$ is classified~\cite{Wen_SPT2} by $\Z_{(2,q)}$ which is $\Z_2$ for even $q$ and trivial for odd $q$. Therefore for odd $q$ the anomaly of the Luttinger liquid considered above automatically vanishes. For even $q$, using standard argument (see for example Ref.~\cite{Lu-Vishwanath}) the Luttinger liquid described in Eq.~\eqref{eq:LuttingerLagrangian}, \eqref{eq:Kmatrix} and \eqref{eq:LuttingerSymm} (with $\Z$ replaced by $\Z_q$ in Eq.~\eqref{eq:LuttingerSymm}) has a nontrivial $\Z_2$ anomaly. The anomaly can be described by the $(2+1)d$ bulk action $(\pi/q) cda$,
where $a\in H^1(\mathcal{M},\Z_q)$ and $c\in H^1(\mathcal{M},\Z_2)$. The anomaly vanishes in the bulk once we re-insist that $a=z$ is in fact a $\Z$-valued (instead of $\Z_q$) gauge field and we recover Eq.~\eqref{zcanomaly} (up to the equivalence relation Eq.~\eqref{eq:equiv}).

The logic is similar for irrational $\nu=p/q$, except that now $\Z$ acts in the low energy theory effectively as a $U(1)$ symmetry (call it $U(1)_z$ to avoid confusing it with the charge conservation $U(1)$). The Luttinger liquid in Eq.~\eqref{eq:LuttingerLagrangian}, \eqref{eq:Kmatrix} and \eqref{eq:LuttingerSymm} (with $\Z$ replaced by $U(1)_z$ in Eq.~\eqref{eq:LuttingerSymm}) now has a t'Hooft anomaly, described by the $(2+1)d$ bulk action $\pi cdA_z$. Again the anomaly vanishes in the bulk when we insist that $A_z=\nu z$ with $z$ being an integer-valued gauge field.

\section{Stability analysis of the $\mathbb{Z}\times \mathbb{Z}$ anomaly}
\label{sec:LLAC}

To further support the nontriviality of the $\mathbb{Z}\times\mathbb{Z}$ anomaly discussed in Sec.~\ref{sec:ZZanomaly}, we perform an explicit stability analysis using Luttinger liquid theory. Specifically, we will show that a symmetric gapped state cannot be achieved through symmetric perturbations using the Haldane's Luttinger liquid stability analysis~\cite{PhysRevX.3.021009,PhysRevB.86.115131}. 

To address the Luttinger liquid physics, we focus on the four low-energy fermionic modes: two right-movers $\psi_{1}$ at $-k_+$, $\psi_{2}$ at $k_+$, and two left-movers $\psi_{3}$ at $-k_-$, $\psi_{4}$ at $k_-$. Via the standard bosonization procedure, we may then describe these low-energy modes in terms of bosons $e^{i \phi_I}$ ($I\in\{1,2,3,4\}$), with the Lagrangian
\begin{align}
\mathcal{L}=-\frac{1}{4\pi}\left[K_{IJ}\partial_t\phi_I\partial_z\phi_J+V_{IJ}\partial_z\phi_I\partial_z\phi_J\right]\quad,
\end{align}
where
\begin{align*}
K=\begin{pmatrix}
1 & 0 & 0 & 0 \\
0 & 1 & 0 & 0 \\
0 & 0 & -1 & 0 \\
0 & 0 & 0 & -1
\end{pmatrix}\quad,
\end{align*}
and $V=v_F\mathds{1}$ is the velocity matrix. 
The fermion creation operators are expressed as $\psi^\dag_I\sim \kappa^+ e^{i\phi_I}$, where $\kappa^+$ are the Klein factors, taking care of the anticommutation relations between different species of fermions. Upon the charge $U(1)$ and translational symmetry transformation, the boson phases change as
\begin{align*}
U(1):&\quad\phi_I\rightarrow \phi_I + \theta\quad,\\
T_z:&\quad\phi_I\rightarrow \phi_I + k_I \quad,
\end{align*}
where $k_I=(-k_+,k_+,-k_-,k_-)$.

We now wish to examine whether we can gap these low-energy modes via symmetry-preserving interactions. A general product of creation and annihilation operators in this language will take the form $e^{i\Lambda^T\phi}$, where $\Lambda$ is a four component integer vector. The particular perturbations we examine are of the form
\beq
U(\Lambda)=U(z)\cos\left(\Lambda^T\phi-\alpha(z) \right)\quad.
\label{eq:Uperturbation}
\eeq
This term describes the scattering of electrons between the modes of opposite chirality. 
Since we have two sets of chiral modes, we must add two sets of backscattering terms $\sum_{i=1}^2 U(\Lambda_i)$ with two linearly independent $\Lambda_1$ and $\Lambda_2$ terms. As the amplitude of $U$ is increased, we eventually will reach a symmetric gapped state, provided there exists a suitable symmetry-respecting set of $\{\Lambda_i\}$. However we will now show that such a set does not exist.

The $\Lambda_i$ must satisfy several conditions, including symmetry conditions
\begin{align}
\label{eq:sym1}
&\sum_I\left[\Lambda_{i}\right]_I=0\quad,\\
&\sum_I\left[\Lambda_{i}\right]_I\,\frac{k_I}{2\pi}= n\quad,
\label{eq:sym2}
\end{align}
derived from the $U(1)$ charge and translational symmetry respectively, where the conditions must be satisfied for some $n\in\mathbb{Z}$. 
We will limit out stability analysis to the charge-conserving Luttinger liquids only, keeping in mind application to the $3+1$d topological semimetals in the 
following section. 

Additionally we must impose the Haldane null vector criterion~\cite{PhysRevX.3.021009}
\beq
\Lambda_i^T K\Lambda_j=0\quad,
\label{eq:Haldane}
\eeq
which essentially guarantees that there exists a linear transformation $\phi\rightarrow M\phi$ that decouples the Lagrangian into two non-chiral Luttinger liquids, which may be gapped out by the backscattering terms. 

We first consider rational values of $k_{\pm}$. Let $\frac{k_+}{2\pi}=\frac{p}{q}$ and $\frac{k_-}{2\pi}=\frac{l}{m}$, where $p,q,l,m\in\mathbb{Z}$ and $p$, $q$ and $l$, $m$ are respective coprimes. The conditions given by Eqs.~\eqref{eq:sym1}, \eqref{eq:sym2} and \eqref{eq:Haldane} yield the general solution
\begin{align}
\Lambda_{i}=\left[\Lambda_{i}\right]_1\begin{pmatrix}
1\\
1\\
-1\\
-1
\end{pmatrix}+\tilde{n}_\pm\begin{pmatrix}
0\\
qm\\
-qm\\
0
\end{pmatrix}\quad,
\label{eq:Lambdasol}
\end{align}
where $\tilde{n}_\pm=\frac{n}{pm\pm lq}\in\mathbb{Z}$. Now that the general solution is found, the final step is to find two linearly independent $\Lambda$ vectors which do not spontaneously break any relevant symmetries when the gap is opened. Specifically we need to check that for all combination $a_1$ and $a_2$ with no common factors there does not exist $a_1\Lambda_1+a_2\Lambda_2=b\Lambda_3$, where $b\in\mathbb{Z}$, such that $\Lambda_3$ is an integer vector that does not obey symmetry constraints given in Eqs.~\eqref{eq:sym1} and \eqref{eq:sym2}. This last condition is known as the \textit{primitivity} condition. It is this constraint in combination with the necessity of two linearly independent $\Lambda_1$ and $\Lambda_2$ terms that causes any solution of the form in Eq.~\eqref{eq:Lambdasol} to fail: any linearly independent choice of $\Lambda_1$ and $\Lambda_2$ will always result in $\Lambda_3=(0,1,-1,0)^T$, which breaks the translational symmetry. Thus we cannot open a gap with perturbations of the form $U(\Lambda)$ without spontaneously breaking any symmetries. A similar consideration for irrational values of $k_{\pm}$ also shows that a gap cannot be opened.

We comment that in principle there is the possibility that a gap can be opened if we include additional ``trivial'' Luttinger liquids into the theory and couple the additional modes with the original modes. Therefore the analysis here does support, but not prove, the nontriviality of the theory.

\section{Lowest Landau Levels for type-I DSM}
\label{app:LLL}

The Hamiltonian of a $C_4$-symmetric type-I Dirac semimetal in 
an external magnetic field along the $z$-direction is given by
\begin{align}
H_\pm(\mathbf{k})&=t \pi_x\sigma^x s^z-t\pi_y\sigma^y+m(0,0,k_z)\sigma^z\quad,
\label{eq:hampm}
\end{align}
where $\pi_{\alpha}=-i\partial_\alpha-A_\alpha$,  $\nabla\times\mathbf{A}=B_z \hat{z}$, and $[\pi_x,\pi_y]=-ieB_z$, $\alpha\in\{x,y\}$. We may easily solve this by squaring the Hamiltonian to give
\begin{align}
H_\pm(\mathbf{k})^2&=t^2\left( \pi_x^2+\pi_y^2-eB_z\sigma^z s^z\right)+ m(0,0,k_z)^2,
\label{eq:hampm2}
\end{align}
which may be written as
\begin{align}
H_\pm(\mathbf{k})^2&=2eB_zt^2\left( a^\dag a +\frac{1}{2}(1-\sigma^z s^z)\right)+ m(0,0,k_z)^2,
\label{eq:hampm2diag}
\end{align}
using the standard ladder operators
\begin{align*}
a=\frac{1}{\sqrt{2eB_z}}\left(\pi_x-i\pi_y\right), && a^\dag=\frac{1}{\sqrt{2eB_z}}\left(\pi_x+i\pi_y\right).
\end{align*}
This gives the LLL dispersions $\pm m(0,0,k_z)$ with the corresponding eigenvectors
\begin{align}
|\Psi_1\rangle=\begin{pmatrix}
1\\
0\\
0\\
0
\end{pmatrix}|0\rangle\quad, && |\Psi_2\rangle=\begin{pmatrix}
0\\
0\\
0\\
1
\end{pmatrix}|0\rangle\quad,
\end{align}
where $|0\rangle$ is the state with $a^\dag a|0\rangle=0$. These are also eigenstates of $C_{4z}$, with eigenvalues $C_{4}\Psi_1=e^{i0}\Psi_1$ and $C_{4}\Psi_2=e^{i\pi}\Psi_2$. 
\end{appendix}

\bibliography{references}

\end{document}